\documentclass[lettersize,journal]{IEEEtran}
\IEEEoverridecommandlockouts

\usepackage[switch]{lineno}
\usepackage{graphicx}  
\usepackage{multirow}
\usepackage{algorithm,algpseudocode}
\usepackage{caption,subcaption}
\usepackage{url}
 
\usepackage{amssymb}
\usepackage{enumitem}
\usepackage{booktabs}
\usepackage{amsmath}
\usepackage{float}
\usepackage{booktabs}
\usepackage{diagbox}
\usepackage{stfloats}
\usepackage{makecell}
\usepackage{tablefootnote}
\usepackage{color}
\usepackage{xcolor}
\definecolor{highlight}{rgb}{0,0,0}

\begin{document}


\title{REMAST: Real-time Emotion-based Music Arrangement with Soft Transition}

\author{
    Zihao Wang${^\dagger}$, 
    Le Ma${^\dagger}$, 
    Chen Zhang, 
    Bo Han,
    Yunfei Xu, 
    Yikai Wang, 
    Xinyi Chen, 
    Haorong Hong, 
    Wenbo Liu, 
    Xinda Wu,
    Kejun~Zhang$^*$, \textit{Member, IEEE},
    \thanks{*Corresponding author: Kejun Zhang(email: zhangkejun@zju.edu.cn) is with Zhejiang University and Innovation Center of Yangtze River Delta, Zhejiang University.}
    \thanks{$\dagger$ Zihao Wang and Le Ma are with the same contribution.}
    \thanks{Zihao Wang, Le Ma, Chen Zhang, Bo Han, Yunfei Xu, Yikai Wang, Xinyi Chen, Haorong Hong, Wenbo Liu, Xinda Wu are with Zhejiang University(email: \{carlwang, maller, zc99, borishan815, wangyik, 3200100845, 3200102545, 3190102475, wuxinda\}@zju.edu.cn).}
    \thanks{Yunfei Xu is with Data \& AI Engineering System, OPPO, Beijing, China(email xuyunfei@oppo.com).}
}

\maketitle
\begin{abstract}
Music as an emotional intervention media has important applications in scenarios such as music therapy, games, and movies. However, music needs real-time arrangement according to changing emotions, bringing challenges to balance emotion real-time fit and soft emotion transition due to the fine-grained and mutable nature of the target emotion. Existing studies mainly focus on achieving emotion real-time fit, while the issue of smooth transition remains understudied, affecting the overall emotional coherence of the music. In this paper, we propose REMAST to address this trade-off. Specifically, we recognize the last timestep's music emotion and fuse it with the current timestep's input emotion. The fused emotion then guides REMAST to generate the music based on the input melody. To adjust music similarity and emotion real-time fit flexibly, we downsample the original melody and feed it into the generation model. Furthermore, we design four music theory features by domain knowledge to enhance emotion information and employ semi-supervised learning to mitigate the subjective bias introduced by manual dataset annotation. According to the evaluation results, REMAST surpasses the state-of-the-art methods in objective and subjective metrics. These results demonstrate that REMAST achieves real-time fit and smooth transition simultaneously, enhancing the coherence of the generated music.
\end{abstract}



\section{INTRODUCTION}
Music as a stimulus media is often used to elicit specific user emotions\cite{koelstra2011deap, drossos2015investigating}, due to the fact that music itself contains a wealth of emotional information\cite{roda2014clustering}.

However, music needs to be arranged in real-time according to changing emotions. Through real-time emotion-based music arrangement, a designated music piece undergoes transformation, resulting in a different composition that expresses certain emotions while preserving similarity with the original music. It enables the manipulation of emotional states based on existing melodies, offering considerable application potential in various scenarios, e.g., music therapy\cite{2012Real}\cite{liu2011real,9175586}, video game soundtracks\cite{doi:10.1080/00140139.2013.825013}\cite{8735930,livingstone2005dynamic}, and movie scores\cite{10.1093/acprof:oso/9780199230143.003.0031}\cite{mcalpine2009approaches}. 

These applications benefit from enhanced immersion\cite{ijerph182312486,2011-10977-00120111001} and emotional regulation\cite{doi:10.1080/00140139.2013.825013}\cite{10.1093/acprof:oso/9780199230143.003.0031} while avoiding discomfort from music switches\cite{doi:10.1080/00140139.2013.825013}\cite{10.1093/jmt/thz013}.

{\color{highlight}
Because of the \textit{fine-grained} and \textit{mutable} nature of the target emotions, real-time emotion-based music arrangement often struggles to balance real-time fit and smooth transition. On one hand, some existing works have attempted to achieve real-time fit by monitoring physiological signals\cite{Kana_MIYAMOTO20222021EDP7171} and user interactions\cite{hadjeres2020anticipation}, yielding promising results. On the other hand, the smooth transition issue has been addressed in areas such as coherent paragraph generation\cite{liang2017recurrent} and facial emotion recognition\cite{NASIR2023109971}, but current real-time emotion-based music arrangement methods\cite{Muhamed2021SymbolicMG}\cite{ Sulun_2022}\cite{ferreira2021learning}\cite{hsiao_tzu_hung_2021_5624519}\cite{Madhok2018SentiMozartMG} have not addressed the smooth transition issue, consequently, disrupting the coherence between different segments of the generated music.
}

\begin{figure}[t]
  \centering
  \includegraphics[width=1.01\linewidth]{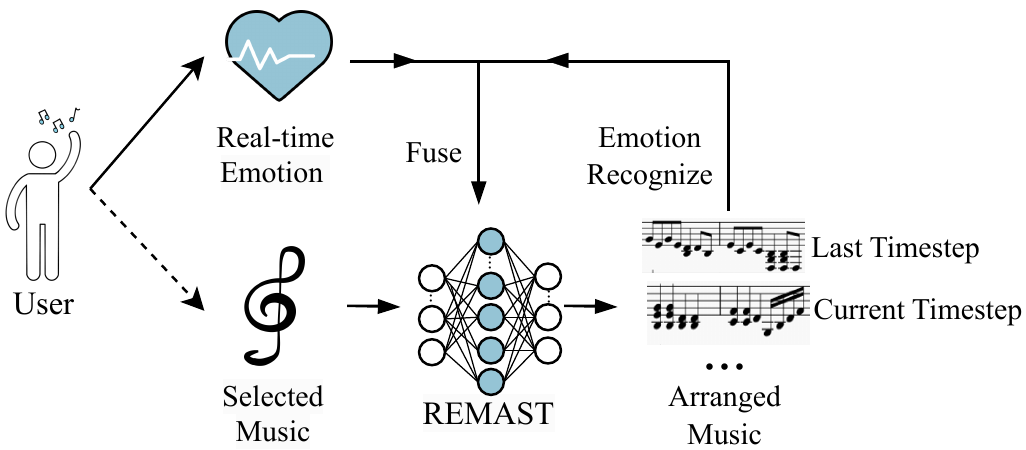}
  \caption{The workflow of utilizing REMAST in various scenarios. The user selects the original music at the beginning, and REMAST obtains the real-time emotion.}
  \label{fig:intro}
\end{figure}

As shown in Figure \ref{fig:intro}, we propose REMAST to balance real-time fit and smooth transition at the same time, consisting of two phases: 1) the music emotion recognition phase, which recognizes the last timestep's music emotion; 2) the music generation phase, which fuses the last timestep's recognized music emotion and current timestep's target input emotion to guide the generation of upcoming melody and harmony.

Specifically, the music emotion recognition model in the first phase is trained in a semi-supervised manner. To better capture the subtle emotion information, we introduce four quantifiable music theory features related to emotion: Harmonic Color\cite{doi:10.1080/00029890.1986.11971924}, Rhythm Pattern\cite{10.3389/fncom.2016.00080}, Contour Factor\cite{Friedmann1985AMF}, and Form Factor\cite{Smith1996MusicalFA}. In the music generation phase, we propose the downsampling arrangement pipeline to tackle the scarcity of emotion-labeled music arrangement data pairs, enabling the adjustment of music similarity and emotion real-time fit by manipulating sampling granularity. Finally, we adopt a texture generation algorithm to transform various harmonies into multi-track accompaniments.

REMAST is evaluated with both objective\cite{yeh2021automatic}\cite{harte2006detecting} and subjective metrics. Experimental results demonstrate that REMAST outperforms state-of-the-art methods\cite{Muhamed2021SymbolicMG}\cite{lucas_ferreira_2019_3527824}\cite{sulun2022symbolic} in music coherence and similarity to the original music while maintaining a high degree of emotion real-time fit, ensuring that REMAST is more suitable for practical applications. {\color{highlight} We also employ REMAST in an anxiety relief application, where emotion-fit and guide play a vital role\cite{schwabe2005resource}. We find that REMAST achieves the best effect in improving participants' emotional state, demonstrating the potential of REMAST in music therapy and other real-world settings.}

In summary, our main contributions are as follows:
\begin{itemize}
    \item We propose REMAST, a real-time emotion-based music arrangement method, which transforms a given music piece into another one to evoke a specific emotion. To the best of our knowledge, we are the first to consider the issue of a smooth transition in real-time emotion-based music arrangement.
    \item We design the arrangement pipeline based on the downsampled original melody, instead of the original melody itself, to tackle the scarcity of data pairs and adjust music similarity and emotion real-time fit flexibly.
    \item We introduce four emotion-related features based on music theory to enhance emotional information.
    \item Both subjective and objective results demonstrate REMAST's effectiveness in enhancing the overall coherence of the arranged music while ensuring emotional real-time fit. 
\end{itemize}


\section{RELATED WORK}

\subsection{Application of Emotional Music}
Emotional music has been applied in many scenarios to evoke specific emotions.

In music therapy, when the music emotion is consistent with the patient's emotion \cite{2011-10977-00120111001}\cite{doi:10.1177/0305735614548500}\cite{ijerph182312486} and the melody is familiar to the patient\cite{Groarke2020Does}\cite{2014Music}, it can effectively promote emotional improvement, trigger patients’ empathy, and thus enhancing therapeutic effects. Starcke et al.\cite{ijerph182312486}suggested that patients should first listen to music that matches their current emotional state, and then listen to music that expresses the desired state, according to the ISO principle. Bower et al. \cite{2014Music}found that using familiar songs as a music therapy intervention can facilitate cognitive recovery after brain injury. Therefore, real-time emotion-based song adaptation can strengthen therapeutic effects by starting from familiar melodies and ensuring real-time consistency between music emotions and patients’ emotions. On the other hand, current music therapy methods based on real-time user emotions, either recommending\cite{2012Real}\cite{liu2011real}\cite{9175586}\cite{app122111209}\cite{Peng2020A}\cite{Zhu2021A}or manually selecting music, are widely used but still suffer from discontinuity caused by changing songs. Sourina\cite{2012Real}used EEG data to identify emotions and recommend songs for therapy, and the system automatically selects the music in the database that matches the user's current emotional state. These methods disrupt the smooth emotional transition when changing songs, causing discomfort and maladaptation for patients\cite{doi:10.1080/00140139.2013.825013}\cite{10.1093/jmt/thz013}\cite{Cai2013The}. Marjolein D. van der Zwaag et al. \cite{doi:10.1080/00140139.2013.825013} found that sudden changes in music increase the sense of sadness. 

Real-time emotion-based music adaptation can enable gradual emotional transitions, avoiding discomfort caused by sudden song changes, which has a wide range of applications in fields such as gaming and film to fit the emotions of each scene to maintain narrative consistency and gradually adjust the audience's emotional state\cite{10.1093/acprof:oso/9780199230143.003.0031}\cite{mcalpine2009approaches}. In addition, real-time emotion-controllable music arrangement has significant practical value in the context of music therapy. Using real-time emotion-controllable music arrangement technology can not only ensure the patient's familiarity with the melody but also achieve a smooth transition of music emotions, thus avoiding the discomfort caused by switching songs for the patient.

\subsection{Controllable Music Generation}
Computer-generated music has become increasingly popular with the development of artificial intelligence. However, most of the music-generation techniques are still uncontrollable or complicated to control, making it difficult for non-music background users to use them. In particular, in music therapy, one of the main functions of music is to arouse the empathy of patients, which requires that the music can fit the emotional state of the patients, so if computer-generated music is to be used in music therapy scenarios, the method needs to be controllable.

Traditional controllable music generation typically uses a rule-based approach. Wallis et al.\cite{rulebased_wallis_2011}\cite{computergenerating_wallis_2008} characterizing the relationship between tempo, articulation, and roughness with varying levels of arousal (the affective dimension). More recently, researchers have used this information to construct more general representations of these rules in parametric equations in which musical features are parameterized in terms of emotion and arousal levels (affective dimensions)\cite{closedloop_ehrlich_2019}. These rule-based methods are governed by a set of mathematical equations and are time and computationally efficient in generating music.

Many optimization methods have also been successfully applied to the automatic creation of emotional music. Scirea et al.\cite{affective_scirea_2017} use a genetic algorithm to select melodies for given chords without violating the rules of music theory. Kuo et al. \cite{development_kuo_2015} designed a tree network for chord selection, where each node in the network is represented by a chord name, and the length of the path between each pair of chords is weighted according to the value of the emotion. The purpose of the cost function is to optimize the total path length by selecting the closest next chord in a sequence of chords with a given sentiment value/emotion value.

Recently, with the deep learning model, data-driven approaches have become the mainstream. Muhamed et al.\cite{Muhamed2021SymbolicMG} combine GAN and Transformer models to generate music guided by authentic music pieces. MuseMorphose\cite{wu2021musemorphose} generates and transforms piano music styles, allowing fine-grained control over attributes such as rhythmic intensity and polyphony. As for emotion-controllable music generation, mLSTM-Ferreira\cite{ferreira2021learning} generates symbolic music based on a given emotion using a generative deep learning model, although its emotion control conditions are limited. SentiMozart\cite{Madhok2018SentiMozartMG} identifies major emotion categories from facial images and generates corresponding music melodies using a two-layer LSTM network. Music Transformer-Sulun\cite{Sulun_2022} proposes controllable music generation based on continuous-valued emotions, conditioning the Transformer model for emotion-controllable music generation. Nonetheless, the current methods predominantly focus on coarse-grained singular control. When applied to real-time music generation tasks, these methods encounter difficulties in preserving overall music coherence under the influence of dynamic fine-grained control. 

\subsection{Real-time Music Generation}
In real-time music generation, RL-Duet\cite{jiang2020rl} employs deep reinforcement learning to predict the next machine note based on previous human and machine music parts. SongDriver\cite{wang2022songdriver} uses a parallel mechanism of prediction and arrangement phases to achieve zero logical latency in real-time accompaniment generation, significantly reducing exposure bias. Robertson et al.\cite{Robertson} propose a method for generating adaptive music in real-time within a virtual environment. 

As for real-time emotion-controllable music generation, Miyamoto et al.\cite{Kana_MIYAMOTO20222021EDP7171} use real-time emotions to generate music and predict changes in the next timestep to avoid logical delays. However, their rule-based music generation method lacks flexibility and richness.

\section{METHOD}

\begin{figure}[t]
  \centering
    \includegraphics[width=.99\linewidth]{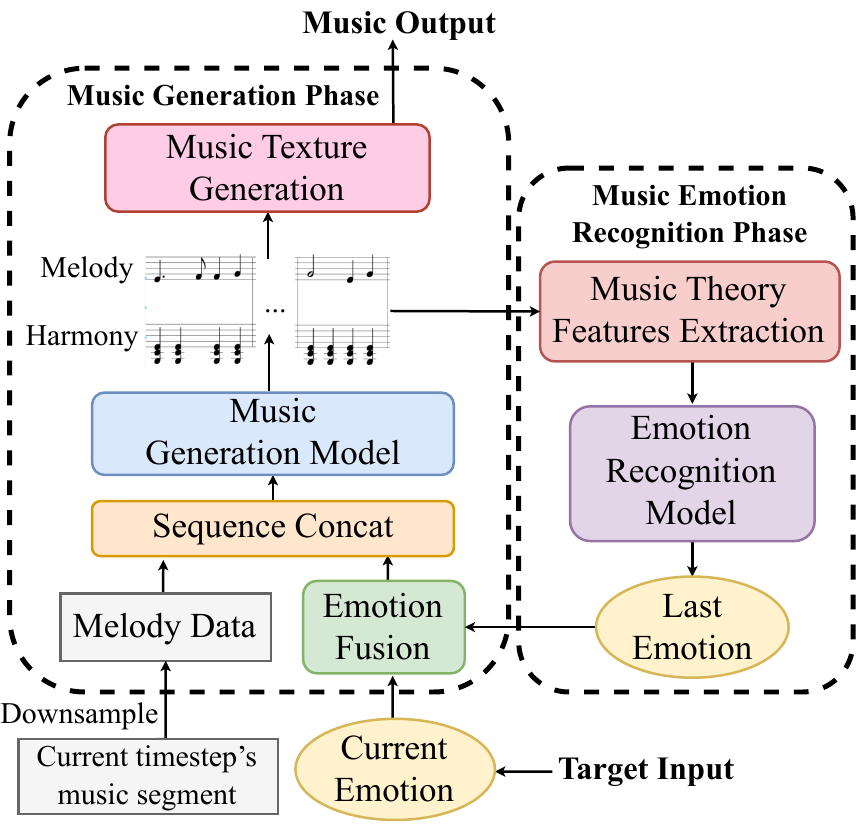}
  \caption{The overall architecture for REMAST. 1) In the recognition phase, REMAST recognizes the emotion of the last timestep's music segment. 2) In the generation phase, REMAST fuses the last timestep's recognized music emotion with the current timestep's target input emotion and generates the current timestep’s music segment based on the fused emotion.}
 \label{figure-architecture}
\end{figure}

\subsection{Overall Architecture}
{\color{highlight} We propose REMAST to generate the melody and chord based on the emotion from the current and last timestep to achieve real-time emotion music generation, where the model scales are limited to reduce the computing latency.}

The structure of REMAST is depicted in Figure \ref{figure-architecture}, which consists of two phases: the music emotion recognition phase and the music generation phase. We fuse the last timestep's recognized music emotion with the current timestep's target input emotion using three different methods. The fused emotion guides REMAST to generate upcoming melody and harmony based on the input melody data. In the recognition phase, REMAST recognizes the emotion of the last timestep's music segment. In the generation phase, REMAST fuses the last timestep's recognized music emotion with the current timestep's target input emotion and generates the current timestep’s music segment based on the fused emotion.

\begin{figure}[t]
  \centering
    \includegraphics[width=0.72\linewidth]{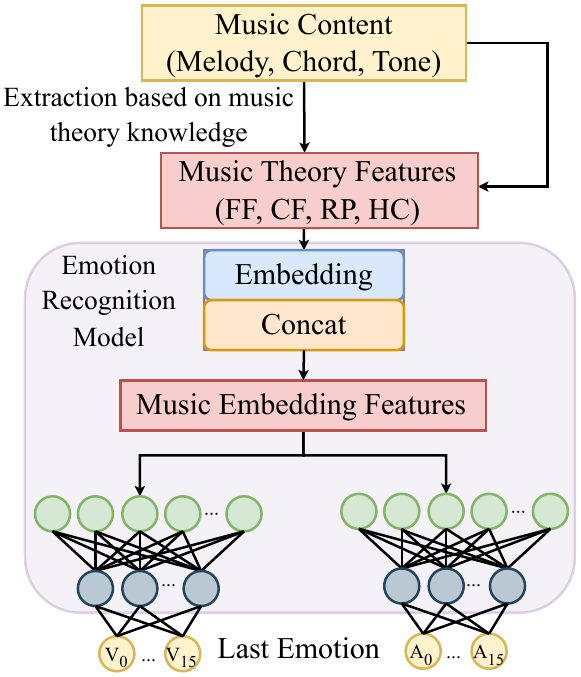}
  \caption{The structure of the music emotion recognition model, which outputs the emotion sequence from the input music content and the music theory features.}
  \label{EmotionModel}
\end{figure}

To achieve real-time recognition of music emotions, we introduce an emotion recognition model utilizing two sets of Multilayer Perceptrons (MLPs) with the structure depicted in Figure \ref{EmotionModel}. The music content and the four music theory features, namely Form Factor (FF), Contour Factor (CF), Rhythm Pattern (RP), and Harmonic Color (HC) are embedded separately and concatenated as the input of the emotion recognition model. Then, the recognition model outputs a fine-grained sequence of Valence-Arousal values corresponding to the four bars of music with granularity consistent with ground-truth emotions.

The music generation model is implemented based on a Transformer. Since the entire melody data is known after the user selects the music, the model's input is the melody of the current timestep, which avoids the issue of logical latency.

\subsection{Music Theory Features Extraction}

To enhance the emotional information of the music segment in the recognition model, we introduce four quantifiable music theory features related to emotion based on music theory and music psychology.

The following music theory features quantify the music's emotion from harmony, melody, musical tension, and musical structure, which are the four aspects of emotional expression in music. 

\begin{figure}[htbp]
  \centering
    \includegraphics[width=0.64\linewidth]{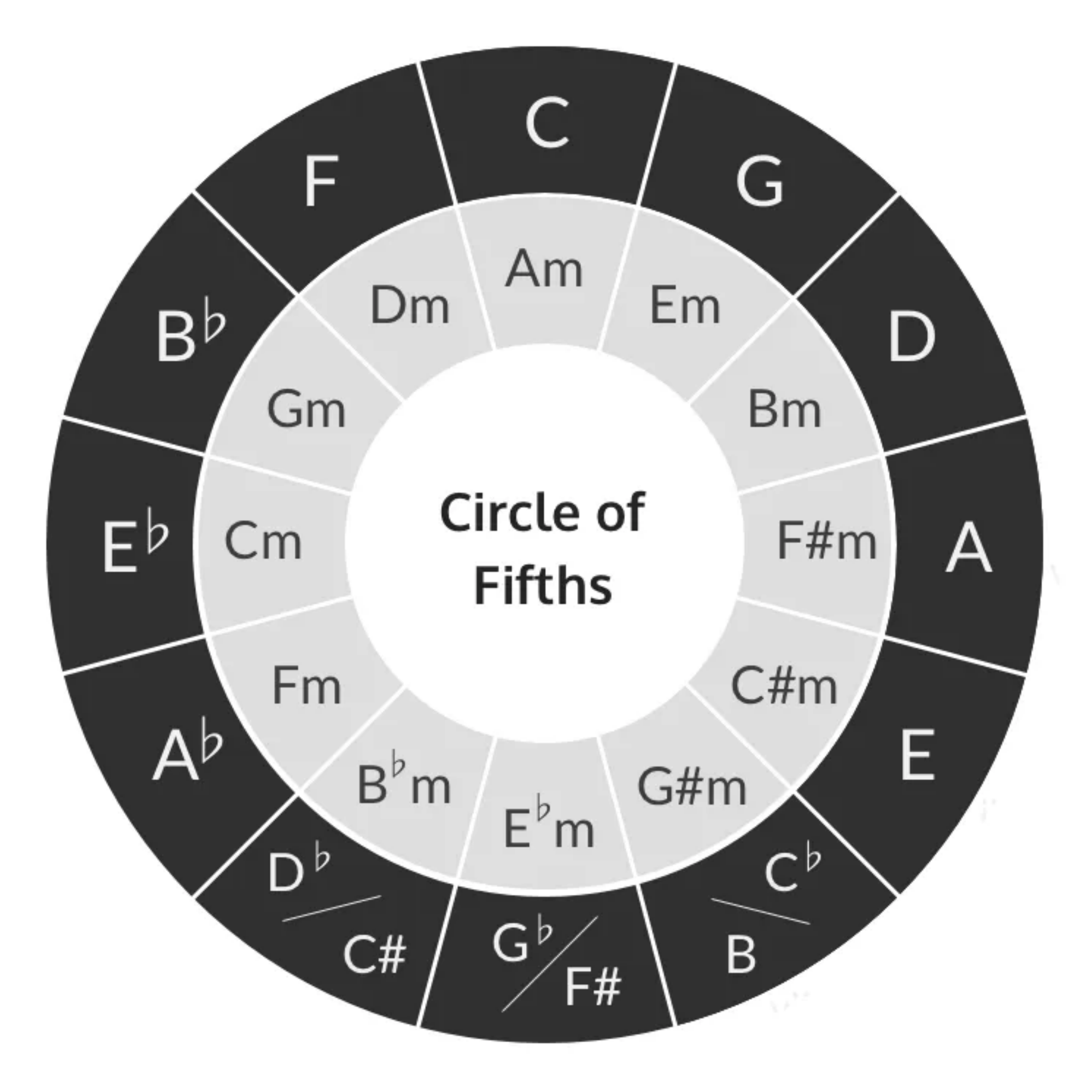}
  \caption{The circle of fifths serves as the foundation for Harmonic Color, quantifying chord freshness by assigning each note a position and value.}
  \label{fig:Chord_Circle}
\end{figure}

\textbf{Harmonic Color} describes the contrasting relationships between different harmonics and alludes to the freshness of harmonic progressions\cite{doi:10.1080/00029890.1986.11971924}. The concept of Harmonic Color is based on the circle of fifths in Figure \ref{fig:Chord_Circle}, We assign 0 for the C position in the outer loop and increase the value by 1 clockwise. For example, the constitution of the C Major chord is C, E, and G. So the values in the circle of fifths for each note in the C Major chord are 0, 4, and 1. The Harmonic Color provides a quantifiable metric for evaluating the relative freshness between two chords.

To compute the relative Harmonic Color value between chord A and chord B, use the following steps:

1) The major triad of the tonality is set as the reference chord and is represented as B. Chord A represents the specific chord on which we want to calculate the Harmonic Color.

2) The K-value between chord A and the reference chord B is calculated, which is obtained as the average of the assigned numbers of the component notes of the specific chord according to the circle of fifths. In (\ref{eq1}), $t$ represents the number of notes in a chord, and $n$ denotes the note's position within the circle of fifths.

\begin{equation} \label{eq1}
K = {{\sum\limits_{i = 1}^{t}n_{i}}/t}
\end{equation}

3) The relative Harmonic Color between chord A and chord B, denoted by $HC_{AB}$, is calculated as (\ref{eq2}). In this formula, $K_{AB}$ represents the difference between the K-values of chords A and B, $n$ is the number of notes in chord A, $m$ is the number of notes in chord B, $a$ denotes a note in chord A, and $b$ represents a note in chord B. The detailed calculation process of this formula is explained as follows: Firstly, compare the K-value of chord A with that of chord B by subtraction and determine the sign of Harmonic Color. Then, calculate the sum of differences among each individual note’s absolute value between chord A and chord B according to the circle of fifths. Finally, normalize the absolute value of the Harmonic Color to ensure that the final result lies within the range of -1 to 1.

\begin{equation}
\begin{aligned}
\label{eq2}
{HC}_{AB} = sgn\left( K_{AB} \right)*norm({\sum\limits_{1 \leq i \leq n, 1 \leq j \leq m,a_{i} \neq b_{t},\forall t}\left| a_{i} - b_{j} \right|})
\end{aligned}
\end{equation}

\begin{figure}[t]
  \centering
    \includegraphics[width=0.99\linewidth]{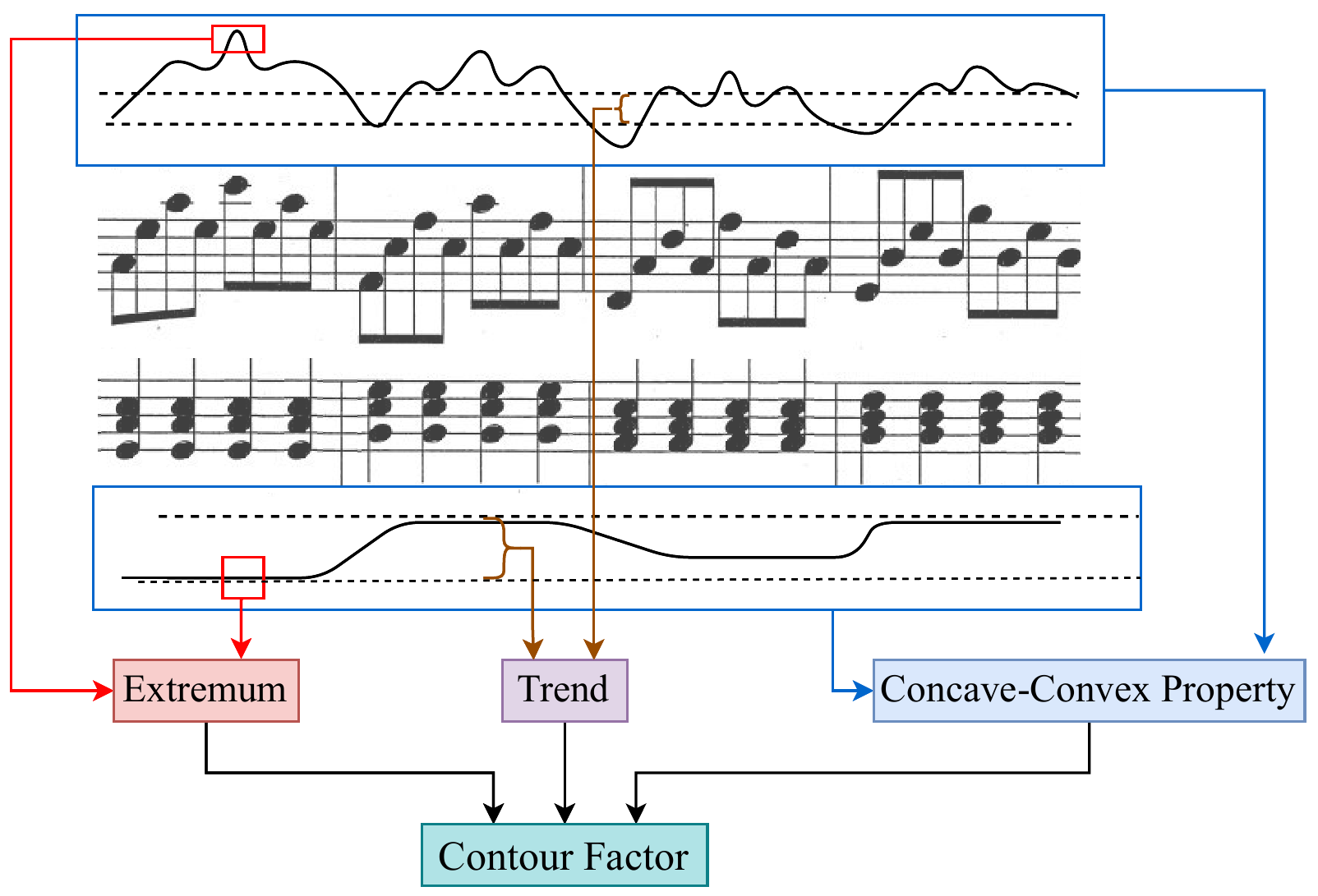}
  \caption{Composition of the Contour Factor, with elements including extremum, trend, concave-convex property.}
  \label{fig:contour_factor}
\end{figure}

\begin{figure*}[htbp]
\centering
\begin{subfigure}{.3\linewidth}
  \centering
  \includegraphics[width=.87\linewidth]{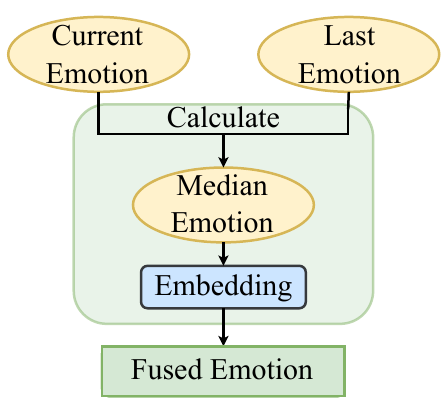}
  \caption{Median Emotion.}
  \label{Median Emotion}
\end{subfigure}
\hfil
\begin{subfigure}{.3\linewidth}
  \centering
  \includegraphics[width=.815\linewidth]{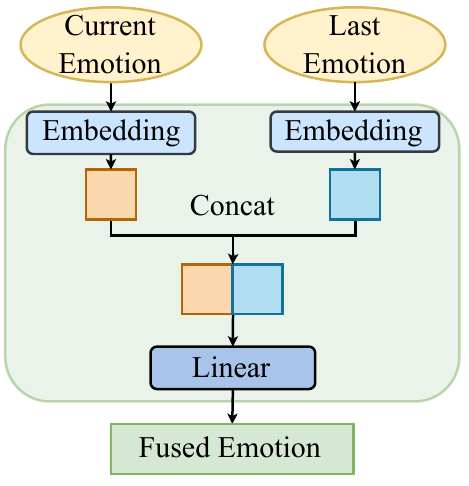}
  \caption{Emotion Concat.}
  \label{Emotion Concat}
\end{subfigure}
\hfil
\begin{subfigure}{.3\linewidth}
  \centering
  \includegraphics[width=.85\linewidth]{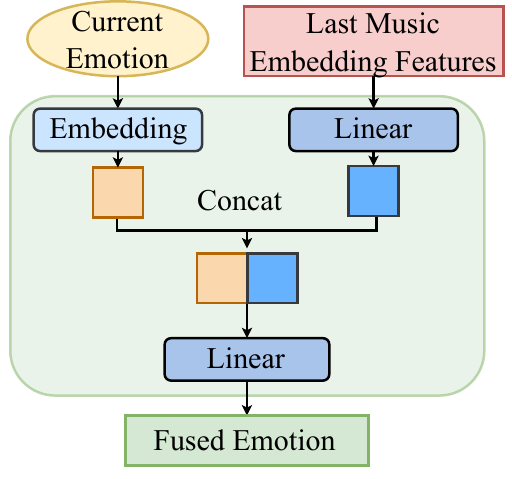}
  \caption{Features Concat.}
  \label{Features Concat}
\end{subfigure}
  \caption{The three different emotion fusion methods. The extraction of music embedding features is described in Figure \ref{EmotionModel}.}
  \label{figure-fusion}
\end{figure*}

\textbf{Contour Factor} reveals the changes of music emotions in time series space and provides a quantitative basis for music tension\cite{Friedmann1985AMF}. As shown in Figure \ref{fig:contour_factor}, the concept of Contour Factor pertains to the variation trends in musical melodies. Specifically, each bar in a piece of music is characterized by three elements that comprise the Contour Factor: the extremum of the melodies, the trend within each four bars, and the concave-convex property of the music shape. All of these concepts are expressed in two dimensions: the lower chord dimension and the higher melody dimension. 

1) To identify the pitch extremums within four bars, the highest pitch in the melody is considered the maximum value, while the lowest pitch among the chord notes constitutes the minimum value, as chords typically serve to accompany and support the melody. 

2) In terms of the trend within each four bars, the chord trend is obtained by calculating the difference between the lowest note of the last chord and the lowest note of the first chord. The melody trend is defined as the pitch difference between the last note and the first note in the melody. 

3) Concerning the concave-convex property, the melody or chord concave-convex property arises from the difference between the average of all pitches in the melody or chord and the mean value of the last and first pitches within four bars. 

The pitch may exhibit a coherent or skipping character as the melody progresses. The combination of different musical contours can make the music complex and delicate, thereby expressing the corresponding emotions more accurately\cite{Friedmann1985AMF}. The extraction of the Contour Factor reveals the changes of music emotions in time series space and provides a quantitative basis for music tension.

\textbf{Form Factor} is the standard format of the melody and harmony from various paragraphs during development\cite{Smith1996MusicalFA}. The Form Factor is a critical set of basic structures that reflects the features of music and ultimately influences the mood of the audience. Five key components of the Form Factor are identified: melody repetition, chord repetition, melody tonality transform, melody zone transform, and melody rhythm difference. As these structural characteristics require relatively global information for reference, we only record all the pitch information of the last 80 bars. As new input is recorded, the earliest information in the queue is removed. This method optimizes memory usage while allowing previously recorded global information to be preserved.

1) Melody repetition is calculated by comparing the current melody segment to the cached melody segments in the queue to determine if there is a pattern of repetition. If the current segment matches one of the cached segments above a certain threshold, the repetition sign is assigned as 1. The interval between these two segments is recorded as the repetition interval. The combination of the repetition sign and the repetition interval constitutes the melody repetition. 


2) Chord repetition closely resembles the method used for Melody Repetition and will not be further elaborated here.


3) Melody tonality transform determines the similarity between melody segments regardless of tonality. Namely, it will be set to 1 if each note of the current melody segment and one of the cached melody segments are in different tonality above a certain threshold of precision, the melody is judged to be melody similarity transform.


4) Melody zone transform is set to 1 when each note of the current melody segment and one of the cached melody segments differs only by some octaves (multiples of 12 in MIDI representation) and their similarity regardless of octaves is above a certain threshold.


5) Melody rhythm difference depends on two conditions. First, the current melody segment and one of the cached melody segments must be similar above a certain threshold, and second, their rhythms must differ significantly under some pre-defined similarity measurement. With all the conditions satisfied, it will be set to 1.

After extracting these five forms of judgment, concatenation of these judgments yields the overall Form Factor. These Form Factors show the differential expression of emotions in the musical structure, such as repetition, contrast, and variation, which provide clear music structural information\cite{Smith1996MusicalFA}.

\textbf{Rhythm Pattern} is an information representation of melody, which reflects the changing law of each note duration in the timing dimension\cite{10.3389/fncom.2016.00080}. The same sequence of notes accompanied by different Rhythm Patterns is bound to bring a new emotional experience. Rhythm Pattern is an information representation of melody that reflects the changing law of each note duration in the timing dimension. Rhythm plays a pivotal role in music mood progression, as different Rhythm Patterns often give rise to varied emotional responses.

To extract Rhythm Patterns from a musical piece, we have adopted an approach that involves recording the duration of each different successive pitch. During the generation process, notes and their corresponding durations can be obtained through the output of the generation model.

It is important to note that when processing the dataset, we extract the Rhythm Pattern in the form of notes and corresponding duration using MIDI information. This step is taken before downsampling the melody and chord into a sparser representation which ensures the completeness of the musical information. This approach helps to maintain the accuracy of the music data extracted from the dataset.

Rhythm pattern is an essential element affecting the music mood, as fast-paced music usually induces positive emotional experiences, such as joy, excitement, and liveliness, while slow-paced music induces negative emotional experiences, such as sadness, depression, and solemnity\cite{10.3389/fncom.2016.00080}.

\begin{figure}[t]
  \centering
    \includegraphics[width=0.75\linewidth]{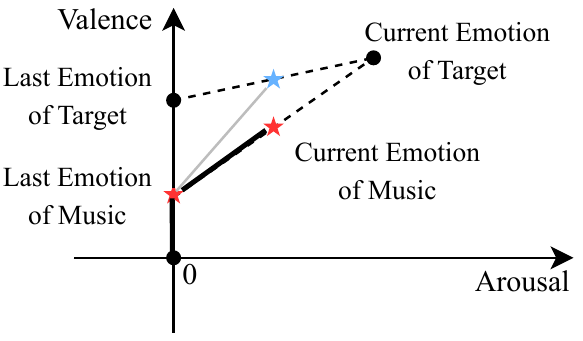}
  \caption{A schematic of the emotion trajectory correction and distance reduction method.}
  \label{fig:emotion_stream}
\end{figure}

\begin{figure}[t]
  \centering
    \includegraphics[width=0.75\linewidth]{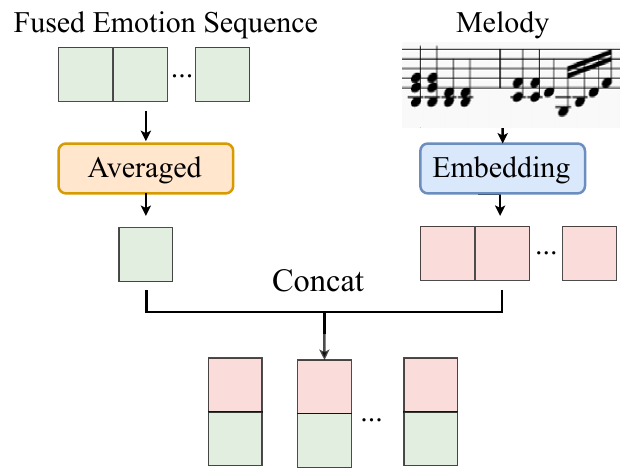}
  \caption{The sequence concat diagram. The tokens depicted at the bottom of the diagram, which incorporate both fused emotion and melody information, will be fed into the music generation model for processing.}
  \label{fig:seq_concat}
\end{figure}

\subsection{Emotion Fusion}
We adopt three emotion fusion methods to integrate emotions from different timesteps, as shown in Figure \ref{figure-fusion}. The justification for fusing emotion based on the last timestep's recognized music emotion and the current timestep's target emotion, rather than the last and current timestep's target emotions, is elaborated in Figure \ref{fig:emotion_stream}. The first red star indicates the previous timestep's generated music emotion position. For the current timestep, two alternative targets, blue and second red stars, are considered. The blue star represents the midpoint of the last and current target emotions, while the second red star denotes the midpoint between the previous music emotion and the current target emotion. By selecting the second red star, the method ensures proper directionality toward the target and reduces emotional distance between timesteps, enhancing the generated music emotion coherence.

Figure \ref{fig:emotion_stream} demonstrates the correction of direction and reduction of distance in the generated music's emotional changes. Compared with fusing the previous and current target emotions, fusing the last music emotion with the present target emotion maintains the consistency of generated music with the user's emotional changes. Additionally, this method shortens the emotional distance between adjacent timesteps, further improving the emotional coherence of the generated music.

\textbf{(a) Median Emotion.}
We compute the median of the last timestep's recognized music emotion and the current timestep's target emotion, which correspond to the midpoints of the emotional values in the Valence-Arousal space. 

\textbf{(b) Emotion Concat.}
After concatenating the emotions of the two timesteps, we use a linear layer to reduce the dimensionality of the features.

\textbf{(c) Features Concat.}
We utilize the music embedding features derived from the music theory features and music content to effectively convey emotional information within the generative model, consequently substituting the last timestep's recognized music emotion. We then concatenate music embedding features with the current timestep's target emotion and reduce the feature dimensionality through a linear layer.

With the fused emotion, we expand the fused emotion feature to the same sequence length as the melody feature and then concatenate them in the sequence dimension. As depicted in Figure \ref{fig:seq_concat}, we perform sequence average on fused emotion sequence as the final emotion representation, which is then repeated and concatenated with each token of the input melody as the fine-grained control condition for music generation.

\subsection{Downsampling Arrangement Pipeline}

Owing to the scarcity of emotion-labeled music arrangement data pairs, we propose the downsampling arrangement pipeline to tackle this challenge by filling in music details. Specifically, drawing inspiration from super-resolution technology in image processing\cite{anwar2020deep}, we first downsample the melody to obtain a low-resolution representation at the beat-level sampling granularity. Next, we generate a high-resolution version (including harmony and melody details, etc.) from the low-resolution representation based on the real-time input emotion. 

The downsampling arrangement pipeline exhibits exceptional flexibility as it enables the adjustment of music similarity and emotion real-time fit by manipulating sampling granularity. For instance, by coarsening the sampling granularity, the similarity of the generated music to the original music can be reduced, while simultaneously enhancing the real-time fit to the target emotion.

\subsection{Semi-supervised Learning}

\begin{figure}[t]
  \centering
  \includegraphics[width=0.95\linewidth]{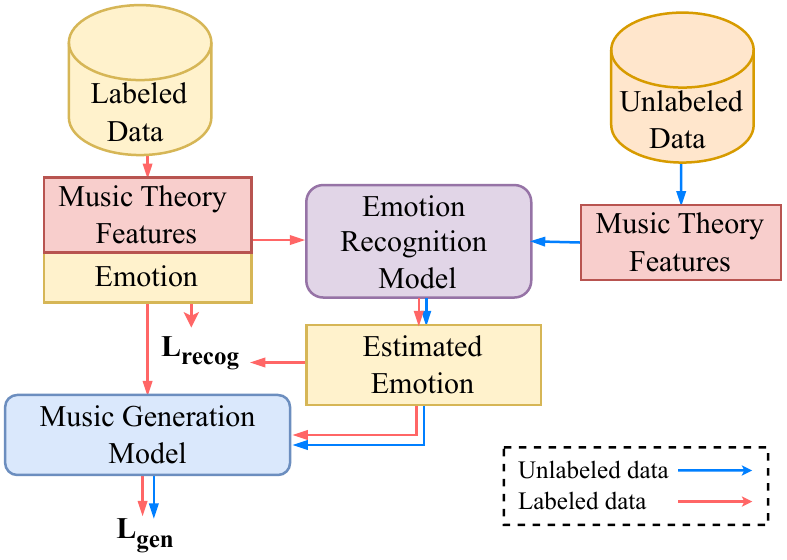}
  \caption{Semi-supervised learning training approach. REMAST is first trained on the emotion-labeled dataset to compute $L_{recog}$ and $L_{gen}$. When converged, REMAST is then trained on the unlabeled dataset only to compute the $L_{gen}$.}
  \label{fig:semi-supervised}
\end{figure}

We introduce semi-supervised learning to make full use of large-scale data with and without emotional labels. As shown in Figure \ref{fig:semi-supervised}, we first train REMAST on the emotion-labeled data with the emotion recognition loss $L_{recog}$ and the music generation loss $L_{gen}$, both of which are calculated through cross-entropy loss between the model output and the ground truth. Then, we train REMAST on the unlabeled data, where the emotion recognition model recognizes the music emotion label for the music generation model. In each iteration, the music generation loss $L_{gen}$ is calculated to update the music generation model and the emotion recognition model.

\begin{equation} \label{eq3}
Emo = (1 - \alpha) * Emo_{label} + \alpha * Emo_{recog}
\end{equation}

\begin{equation} \label{eq4}
\alpha = \frac{N_{cur} }{ N_{total}}
\end{equation}

Besides, during semi-supervised learning, We reduce subjective bias introduced by manual dataset annotation by integrating prior knowledge from the emotion recognition model. Specifically, we integrate the Valence-Arousal values from both the emotion-labeled dataset and the emotion recognition model to train the music generation model. Utilizing the generated music as a supervised signal, we optimize the recognition model, consequently reducing the impact of bias introduced by manual dataset annotations. The method is shown in the (\ref{eq3}).

\section{DATASET}

\begin{table*}[htbp]
  \caption{Statistics of eleven open-source music datasets. (TVE: Time-Various Emotions; GE: General Emotions; VAR: Valence-Arousal Ranges)}
  \centering
  \resizebox{0.9\textwidth}{!}{%
    \begin{tabular}{ccccccc}
      \toprule 
      Datasets & \#clips & TVE & GE & VAR & File Format & {\color{highlight}Styles/Genres}\\
      \midrule
      Theorytab\tablefootnote{https://www.hooktheory.com/theorytab} & 11270 & - & - & - & xml & All modern music genres \\
      Wikifornia\tablefootnote{http://www.wikifonia.org} & 4017 & - & - & - & musicxml & All modern music genres \\
      Nottingham\cite{boulanger2012modeling} & 591 & - & - & - & midi & American and British folk songs \\
      àiSong\cite{wang2022songdriver} & 2323 & - & - & - & text & Chinese-style modern pop music \\
      \midrule
      PMEmo\cite{zhang2018pmemo} & 2881 & \checkmark & \checkmark & [0, 1] & audio(voice) & Rock, classical, pop, electronic \\
      EmoMusic\cite{soleymani20131000} & 704 & \checkmark & $\times$ & [-1, 1] & audio(voice) & Rock, classical, pop, electronic \\
      DEAM\cite{soleymani2016deam} & 1680 & \checkmark & \checkmark & [-1, 1] & audio(voice) & All modern music genres \\
      VGMIDI\cite{ferreira2021learning} & 3099 & \checkmark & $\times$ & [-1, 1] & midi & Video game soundtracks \\
      C-WCMED\cite{fan2020comparative} & 492 & $\times$ & \checkmark & [-1, 1] & audio & Western and Chinese classical music \\
      EMOPIA\cite{hsiao_tzu_hung_2021_5624519} & 1078 & $\times$ & \checkmark & 4 Quadrants & midi & Pop piano music \\
      Soundtracks\cite{Eerola2011ACO} & 710 & $\times$ & \checkmark & 3-Dim & audio & Film soundtracks \\
      \bottomrule
    \end{tabular}%
  }
  \label{tabdataset}
\end{table*}

We leverage eleven open-source music datasets, comprising seven emotion-labeled datasets and four unlabeled datasets. The statistics of the datasets are shown in Table \ref{tabdataset}. {\color{highlight}The huge amount of data covers nearly all music genres, guaranteeing the model to generalize to different music genres and emotional contexts.}

\subsection{Data Processing}

To address inconsistencies in file formats and Valence-Arousal ranges across datasets, we first perform several processing steps as follows:

\textbf{Audio-MIDI Format Conversion.} , we transcribe them into symbolic data. For audio data of pure musical instruments, we directly use the Onsets $\&$ Frames method\cite{hawthorne2018onsets} to transform them into MIDI format. For audio data containing vocal voice, we first use the Onsets $\&$ Frames method\cite{hawthorne2018onsets} to identify midi and place it on the second track, and then use the world vocoder in the Harvest method\cite{morise2017harvest} to identify the pitch corresponding to the voice frequency and put it on the first track, finally, remove the duplicate notes in the second track and the first track.

\textbf{Align Label with Content.} We adjust and align the positions of emotion tags and music content according to the time of emotion tags and the BPM of midi.

\textbf{Data Screen.} For all datasets, only music pieces with time signatures of 4/4 and 2/4 have been preserved for subsequent sampling. We also remove the end clips of less than 4 bars. We screen the transcribed data by removing those with poor transcription effects. The \#clips shown in Paper, Table 1 reflects the outcome of our rigorous data processing and filtering procedures. 

Then, we cut MIDI files into pieces of data with a length of 4 bars and perform the following steps:

\begin{figure}[t]
  \centering
    \includegraphics[width=1.03\linewidth]{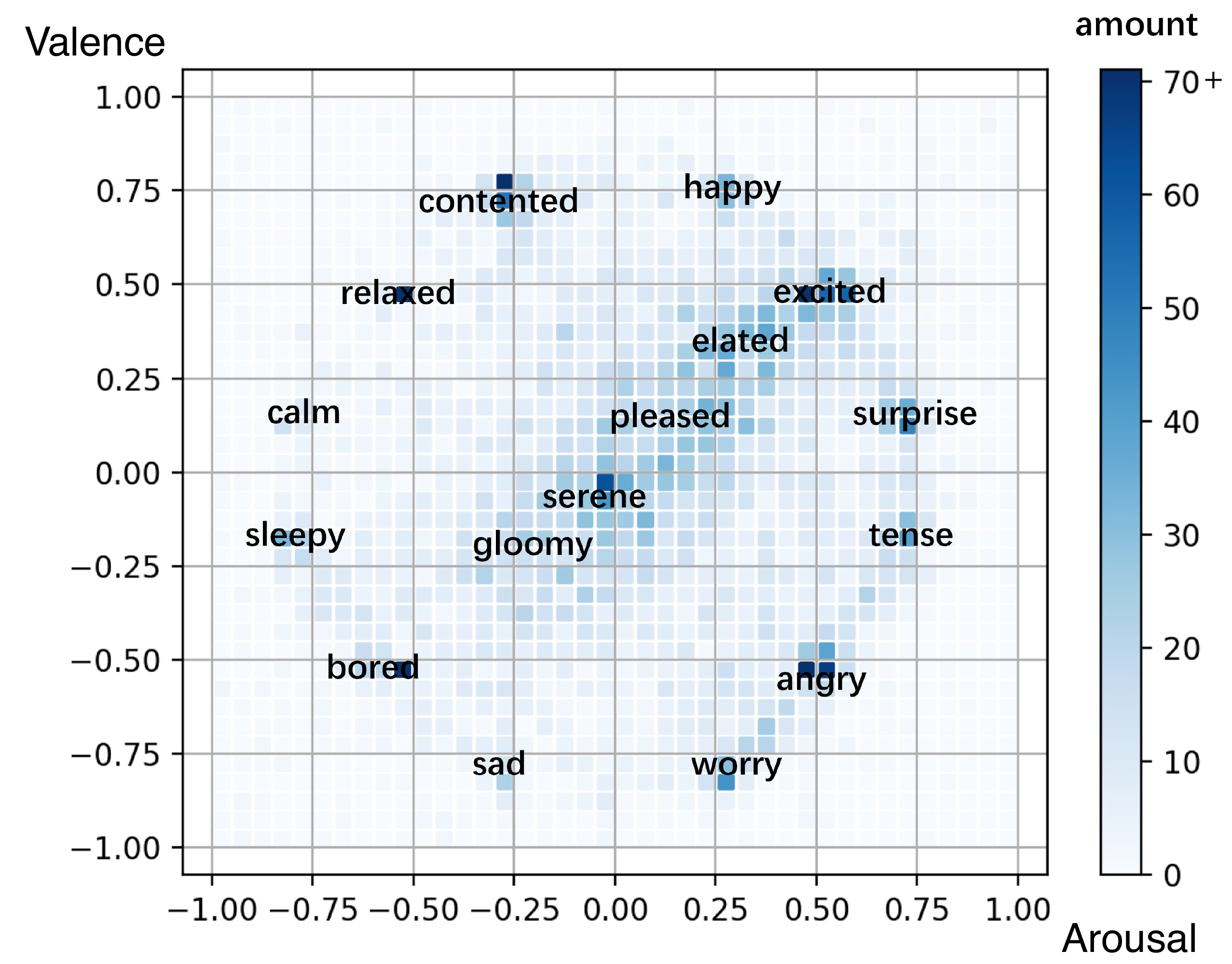}
  \caption{Distribution of datasets in the V-A space and mapping of discrete emotion labels to V-A values.}
  \label{fig:emo_map}
\end{figure}

\textbf{Melody Symbol Extraction.} For each piece of data, we quantify and round the duration of each note in the melody track of the MIDI files in units of the sixteenth note to obtain the melody sequence. Simultaneously, we downsample the melody, sampling the pitch of the melody every quarter note, and obtain the downsampled melody sequence that will be used for Downsampling in the Arrangement Method.

\textbf{Harmony Symbol Extraction.} For harmony extraction, the datasets with emotional labels divide the multi-track MIDI files into quarter notes as a segment and automatically analyze the notes in each segment to obtain its harmony label and then obtain the entire Harmony sequence. To ensure consistent harmony sampling granularity, we delete music pieces with harmony annotations shorter than a quarter note from emotionless datasets that already contain pre-annotated harmony information.

For discrete emotion, we adopt Russell's circumplex model of affect as our basis\cite{russell1980circumplex}\cite{kim2020situ}, mapping 16 discrete emotional states onto the continuous valence-arousal (V-A) space, thus establishing a connection between them. Specifically, we analyze the distribution of emotion annotations in the V-A space within the emotion-music dataset. As depicted in Figure \ref{fig:emo_map}, the color blocks corresponding to different emotions represent a normal distribution centered around the respective emotion's V-A values, while the shades of color reflect the amount of emotion data within the distribution. Then we select 12 uniformly distributed emotions from the circumferential, data-dense regions. Additionally, we introduce four emotions distributed within the data-dense regions, ultimately achieving a mapping of 16 discrete emotion labels to V-A values.

Eventually, to unify the emotion, aiming at the energy arousal and tension arousal three-dimensional problem in the Soundtracks dataset, we eliminate one of the two highly correlated dimensions (tension arousal and valence) and convert them into a two-dimensional VA format with linearly mapping methods\cite{Eerola2011ACO}. For the EMOPIA dataset, we map the data item (4Q format) into four normal distribution spaces with (0.5,0.5) (0.5,-0.5) (-0.5,0.5) (-0.5,-0.5) as the center point separately. Besides, to resolve the issue of varying Valence-Arousal ranges in datasets with emotional labels, we apply Min-Max normalization to change the V-A value range in the PMEmo dataset from [0, 1] to [-1, 1].

\subsection{Data Representation}
Data pieces consist of tonality, melody sequence, downsampled melody sequence, and harmony sequence. For emotion-based datasets, Valence-Arousal values are also included in the data pieces.

Through the aforementioned steps, we obtain 18,201 labeled data pieces and 15,591 unlabeled data pieces for subsequent emotional music generation tasks.

Each data piece contains tonality, melody sequence, downsampled melody sequence, and harmony sequence. Additionally, each data piece in the emotion-based music datasets includes Valence-Arousal values. The melody sequence represents the pitch obtained by sampling every sixteenth note, while the downsampled melody sequence represents the pitch obtained by sampling every quarter note. The harmony sequence represents the harmony obtained by sampling the accompaniment every quarter note, with each harmony consisting of 1 to 5 notes.

\section{EXPERIMENTS}

\subsection{Experimental Setup}

REMAST is implemented using PyTorch, with the recognition model based on two sets of MLPs and the generation model based on a Transformer. The hidden layer size of the recognition model is set to 512, and ReLU is used as the activation function. For the generation model, the maximum input length is 64, the maximum output length is 256, the embedding dimension is 512, the feedforward network dimension is 1024, the number of attention heads is 2, and the dropout is set to 0.1. Both the Transformer Encoder and Decoder have 4 layers.

In the experiments, the dataset is divided into 80$\%$ for training, 10$\%$ for testing, and the remaining 10$\%$ for validation. In the training process, we use the Adam optimizer with a learning rate of 1e-4 and a batch size of 128. The model is trained for 50 epochs on an NVIDIA Tesla A100. 

We combine the melody data of popular songs with fine-grained dynamic emotion sequences and obtain 180 pieces of results as the test input, each with a total length of 60 bars.

\subsection{Evaluation Metrics}

\subsubsection{Objective Evaluation Metrics}
For objective evaluation, we use the following metrics to evaluate REMAST from coherence, similarity between original and generated music, and the real-time fit degree:

\textbf{1) PCC:} The concept of Pitch Consonance Coherence (PCC)\cite{yeh2021automatic} aims to quantify the consonance coherence between different music segments. By utilizing Pitch Consonance Score (PCS) which represents the tonal interval differences between the melody and the associated chord, we compute the PCC by calculating the difference between the PCS values of the last and the current music segments.

\textbf{2) CEC:} The concept of Chord Entropy Coherence (CEC)\cite{yeh2021automatic} aims to quantify the coherence in chord richness and emotional intensity across music segments. By utilizing Chord Histogram Entropy (CHE) which measures the entropy of a given chord sequence and reaches its maximum when the chord sequence follows a uniform distribution, we compute the CEC by calculating the difference between the CHE values of the last and the current music segments.

\textbf{3) MCTC:} The concept of Melody-Chord Tonal Coherence (MCTC)\cite{harte2006detecting} aims to quantify the coherence in harmony and emotional pleasantness between music segments. By utilizing Melody-Chord Tonal Distance (MCTD) which calculates the Euclidean distance between the melody vectors and the chord vectors in a 6D linear space, we compute the MCTC by calculating the difference between the MCTD values of the last and the current music segments.

\textbf{4) overall coherence:} Considering that higher values in the three metrics mentioned above indicate poorer coherence, we subtract their sum from a fixed value to make the value consistent with the quality. This new metric reflects the overall coherence of the music in various aspects.

\textbf{5) similarity:} To ensure a certain degree of similarity after arrangement, we assess pitch similarity between the arranged and original music.

\textbf{6) real-time fit:} We calculate the Euclidean distance between the Valence-Arousal values of music emotion and target emotion at the current timestep to explore the emotion real-time fit degree. Since a higher distance indicates a poorer fit, we subtract it from a fixed value to make the value consistent with the quality.

\subsubsection{Subjective Evaluation Metrics \& Participants}

For the subjective evaluation, we designed a survey questionnaire {\color{highlight}and invited thirty participants (12 women, 18 men) from different backgrounds to minimize the effect of subjective bias and individual preferences. The participants include fifteen professionals with three years' average music performance experience and fifteen amateurs.}

In the experiment, participants first choose popular songs they are familiar with and fine-grained emotion sequences. We then use different methods to arrange the original music by the emotion sequence. Participants listen to the original music containing melody and harmony first, followed by the model-arranged versions. They are asked to observe and judge the softness of the music transitions during abrupt emotion changes, the fit of the arranged music to the target emotion sequences, the similarity to the original music, and the overall coherence of the music.

\begin{figure}[]
  \centering
    \includegraphics[width=1.03\linewidth]{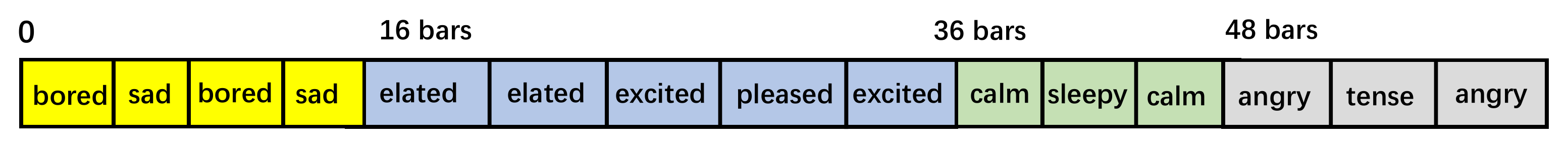}
  \caption{Emotion variation bar chart. The same color represents similar emotions, while different colors represent different types of emotions.}
  \label{fig:exp_color_bar}
\end{figure}

\begin{table*}[t]
  \small
 \centering
  \caption{Objective results of REMAST and baseline methods.}
     \begin{tabular}{cccccccc}
      \toprule
      \multirow{3}{*}{Methods} & \multicolumn{6}{c}{Objective Metrics}   \\
       \cmidrule(lr){2-7} 
         &  \multicolumn{4}{c}{coherence} & \multirow{2}{*}{similarity~$\uparrow$ } & \multirow{2}{*}{real-time fit~$\uparrow$ }	  \\
       \cmidrule(lr){2-5}
        & PCC ~\cite{yeh2021automatic}$\downarrow$ & CEC~\cite{yeh2021automatic}$\downarrow$ & MCTC ~\cite{harte2006detecting}$\downarrow$ & overall~$\uparrow$ & &   \\
        \midrule
        TG-Muhamed\cite{Muhamed2021SymbolicMG} ~ & 3.62±0.29 & 7.51±0.66 & 1.77±0.13 & 1.10±0.72 & 6.67±0.03 & \textbf{2.08±0.71} \\
        mL-Ferreira\cite{lucas_ferreira_2019_3527824}~ & 3.12±0.13 & 5.09±0.29 & 1.35±0.07 & 4.44±0.32 & 6.91±0.01 & 1.58±0.82 \\
        MT-Sulun\cite{sulun2022symbolic}~ & 3.38±0.27 & 4.57±0.30 & 1.73±0.16 & 4.32±0.42 & 6.11±0.04 & 1.67±0.08 \\ 
        \midrule
           REMAST & \textbf{3.04±0.19} & \textbf{3.71±0.31} & \textbf{1.04±0.09} & \textbf{6.21±0.37} & \textbf{7.60±0.59} & 2.02±0.74  \\
        \bottomrule
     \end{tabular}
 \label{tab:cmp_objective} 
\end{table*}

\begin{table*}[t]
\small
\centering
\caption{{\color{highlight}Subjective results of REMAST and baseline methods. (All subjective metrics exhibit statistically significant differences with $p < 0.03$.)}}
    \begin{tabular}{ccccccc}
    \toprule
    \multirow{2}{*}{Methods} & \multicolumn{5}{c}{Subjective Metrics}  \\
    \cmidrule(lr){2-6}
    & coherence & softness & similarity & real-time fit	& overall  \\
    \midrule
    TG-Muhamed\cite{Muhamed2021SymbolicMG} & 3.34 & 3.03 & 3.31 & 3.07 & 12.76\\
    mL-Ferreira\cite{lucas_ferreira_2019_3527824} & 3.86 & 3.31 & 3.76 & 3.10 & 14.03\\
    MT-Sulun\cite{sulun2022symbolic} & 2.66 & 2.66 & 2.93 & 2.55 & 10.79 \\   
    \midrule
    REMAST & \textbf{4.00} & \textbf{3.79} & \textbf{3.90} & \textbf{3.66} & \textbf{15.34} \\
    \bottomrule
     \end{tabular}
 \label{tab:cmp_subjective} 
\end{table*}

As participants may have difficulty understanding the specific meanings and differences of the Valence-Arousal values adopted by REMAST, we transform the emotional Valence-Arousal values into discrete emotion labels based on Russell's circumplex model of affect theory and create emotion variation bar charts for participants to reference \cite{russell1980circumplex}\cite{kim2020situ}. As shown in Figure \ref{fig:exp_color_bar}, we use color changes in the emotion variation bar chart provided to participants to display the time positions of each emotion mutation, allowing users to focus on observing the music changes at those time points. For every four-bar emotional sequence, we summarize a discrete emotion label as a prompt. Participants can compare the emotion variation bar chart while listening to the music.

We employ four different subjective metrics and use t-tests for statistical data comparison:

\textbf{1) coherence:} The overall coherence of the music between the former and latter segments in terms of listening experience.

\textbf{2) softness:} The softness and coherence of music emotion transition when significant changes occur in the target emotion sequences.

\textbf{3) similarity:} The similarity between the arranged and original music.

\textbf{4) real-time fit:} The consistency and synchronization between the target emotion and the music emotion of the corresponding timestep.

\subsection{Objective Evaluation}
We compare the performance of REMAST with state-of-the-art baseline methods for conditionally controlled music generation (Transformer-GANs-Muhamed (TG-Muhamed)\cite{Muhamed2021SymbolicMG}), emotion-controlled music generation (mLSTM-Ferreira (mL-Ferreira)\cite{lucas_ferreira_2019_3527824}), and emotion-controlled music arrangement (Music Transformer-Sulun (MT-Sulun)\cite{sulun2022symbolic}) on the merged dataset.

We maintain the original architecture of these three SOTA methods and modify them for real-time emotion-based music arrangement tasks. Specifically, we modify their inputs to be unarranged melodies and outputs to be melodies and harmonies corresponding to the target emotion sequences. We also standardize the length of their input and output sequences to be the same as REMAST.

Considering the results in Table \ref{tab:cmp_objective}, REMAST maintains a high degree of emotion real-time fit while possessing higher music coherence and similarity. Consequently, REMAST has achieved a balance between real-time fit and smooth transition, making it more suitable for meeting human aesthetic needs and holding more excellent practical application value. 

REMAST not only outperforms the other three methods in overall coherence, but also excels in the three sub-metrics of music coherence: PCC, CEC, and MCTC. This demonstrates that REMAST generates more coherent arranged music by integrating the last timestep's emotional features into the current timestep's target input emotion.

REMAST performs best in similarity, indicating that its downsampling arrangement pipeline can preserve the fundamental features of the original music while generating music details, thus achieving higher similarity.

REMAST ranks second on the emotion real-time fit, just behind TG-Muhamed. This may be due to the fact that TG-Muhamed's discriminator judges the relationship between generated and real music, allowing TG-Muhamed to capture target emotion-related features and generate music that better fits the target emotion. However, considering that REMAST maintains relatively high emotion real-time fit while significantly outperforming TG-Muhamed in music coherence and similarity to the original music, REMAST remains the best method.

\begin{table*}[t]
  \small
 \centering
 \caption{{\color{highlight}The objective and subjective analysis of combinations of arrangement pipelines and emotion fusion methods. Setting \#1 excels in three subjective metrics (overall, real-time fit, similarity) with $p<0.012$.}}
 \resizebox{\linewidth}{!}{
 \begin{tabular}{cccccccccccccc}
  \toprule
\multirow{3}{*}{\makecell{Arrangment\\Pipelines}}& \multirow{3}{*}{Setting} &\multirow{3}{*}{\makecell{Emotion\\Fusion\\Methods}}& \multicolumn{6}{c}{Objective Metrics} & \multicolumn{5}{c}{{\color{highlight}Subjective Metrics}}  \\
   \cmidrule(lr){4-9} \cmidrule(lr){10-14}
    & &  & \multicolumn{4}{c}{coherence} & \multirow{2}{*}{similarity$\uparrow$} & \multirow{2}{*}{real-time fit$\uparrow$}	& \multirow{2}{*}{coherence} & \multirow{2}{*}{softness}	& \multirow{2}{*}{similarity} & \multirow{2}{*}{real-time fit}	& \multirow{2}{*}{overall}  \\
   \cmidrule(lr){4-7} 
    & & & PCC\cite{yeh2021automatic}$\downarrow$& CEC\cite{yeh2021automatic}$\downarrow$& MCTC\cite{harte2006detecting}$\downarrow$& overall$\uparrow$& &  &  &  &  &  &  \\
  \midrule
        \multirow{3}{*}{Downsampling} & \#1 & Feature Concat~ & \textbf{3.04±0.19} & \textbf{3.71±0.31} &  1.04±0.09 & \textbf{6.21±0.37} & \textbf{7.60±0.59} & 2.02±0.74  & \textbf{4.00} & 3.79 & \textbf{3.90} & \textbf{3.66} & \textbf{15.34}  \\ 
  
  & \#2 & Median Emotion ~ & 3.34±0.19 & 3.89±0.38 &	1.06±0.15  & 5.71±0.43 & 7.59±0.54 & 2.13±0.82  & 3.93 & \textbf{3.93} & 3.83 & 2.97 & 14.86 \\

       & \#3 & Emotion Concat~ & 3.35±0.19  & 3.87±0.48	& \textbf{0.97±0.15}  & 5.81±0.52 & 7.54±0.60 & 2.03±0.71  & 3.45 & 3.17 & 3.03 & 3.03 & 12.69 \\ 
  \midrule
  \multirow{3}{*}{\makecell{w/o \\ Downsampling}} & \#4 & Feature Concat~ & 3.32±0.21 & 3.92±0.32 & 1.32±0.14 & 5.44±0.39 & 6.48±0.46 & 2.16±0.73 & 3.38 & 3.31 & 2.17 & 2.79 & 11.66\\ 

  & \#5 & Median Emotion~ & 3.67±0.22 & 4.73±0.39 & 1.31±0.17 & 4.29±0.47 & 6.40±0.52 & 2.12±0.66 & 3.52 & 3.41 & 2.10 & 2.90 & 11.93  \\ 

 & \#6 & Emotion Concat~ & 3.62±0.22 & 3.88±0.28 & 1.33±0.16 & 5.17±0.35 & 6.27±0.52 & \textbf{2.20±0.65} & 3.55 & 3.45 & 1.97 & 2.90 & 11.86\\ 

        \bottomrule
 \end{tabular}
 }
 \label{tab:combine_results} 
\end{table*}

\subsection{Subjective Evaluation}

Table \ref{tab:cmp_subjective} presents the average opinion scores for the subjective metrics. The results demonstrate the effectiveness of our method: REMAST significantly outperforms the other methods in all metrics (with $p<0.03$ for each one). This further demonstrates that REMAST is the best method for this task.

The coherence scores indicate that dynamic changes in emotional conditions may compromise other methods' music coherence. In contrast, REMAST generates music with higher coherence in terms of listening experience, demonstrating its ability to achieve smooth transition when faced with dynamic changes.

Based on the softness scores and participants' feedback, REMAST-generated arranged music exhibits better stability at positions where emotions undergo abrupt changes compared to other methods. Additionally, REMAST achieves the highest score in the similarity to the original music.

We notice subtle differences between the objective and subjective evaluation results for real-time fit. After consulting with participants, we found that the harmony of music in terms of listening experience can affect people's emotional perception of the music. This led to REMAST receiving a higher emotion real-time fit score in the subjective evaluation. 

We analyze the differences in calculating the fitting degree between subjective and objective metrics. The objective real-time fit metric is based on the statistical method, reflecting the L2 distance between the emotions detected by the recognition model and the target emotions in various music segments. On the other hand, subjective metrics vary among individuals, as different people perceive different accented sounds when listening to the same piece of music, as shown by research conducted by Rasmus Bååth\cite{10.1525/mp.2015.33.2.244}. 

However, models with high objective real-time fit metrics may contain sudden changes in the generated music sequence, causing a sense of fragmentation between segments and making it difficult for listeners to appreciate the emotions expressed in each segment\cite{doi:10.1080/00140139.2013.825013}\cite{10.1093/jmt/thz013}\cite{Cai2013The}. This ultimately results in a lower perceived fitting degree in subjective experiments. This is also the reason for TG-Muhamed's relatively low target emotional-fitting score in subjective metrics, although the objective metrics score higher.

In contrast, due to the improvement of auditory perception, music with good coherence can enhance its emotional expression ability and enhance real-time emotional fitting, leading to higher emotional fitting in subjective evaluation for REMAST.

\subsection{Effect of Downsampling}

Drawing inspiration from contrastive learning\cite{1640964}, in addition to the downsampling arrangement pipeline, we also investigate the w/o downsampling arrangement pipeline, which involves constructing positive samples by applying noise masking, duration stretching and contraction, key transposition, and sound zone transposition to randomly selected segments of a song.

To select the best combination of music arrangement pipelines and emotion fusion methods for REMAST, we train and compare six different combinations and assess the influence on the generated music. We also conduct objective and subjective evaluations. All metrics and settings are the same with comparison experiments.

As shown in Table \ref{tab:combine_results}, in objective evaluation results, the combination of the downsampling arrangement pipeline and the Features Concat emotion fusion method (Setting \#1) outperforms all other combinations in overall coherence, similarity, and two coherence sub-metrics (PCC and CEC). Although Setting \#1 is slightly outperformed by Setting \#3 in MCTC, the significantly higher overall coherence score for Setting \#1 still makes it the best combination in music coherence. This may be attributed to the downsampling arrangement pipeline preserves the basic structural features of the input original music to help focus on the overall musical structure and enhance the arranged music's coherence and similarity to the original music. Nevertheless, Setting \#6 has the best score in real-time fit, and this may be due to the w/o downsampling arrangement pipeline can generate music with more notes and different patterns, resulting in better real-time fit. Considering all metrics, Setting \#1 is the best method combination.

{
\color{highlight}
Considering the subjective metric analysis in Table \ref{tab:combine_results}, all the downsampling methods obtain slightly better performance than w/o downsampling methods and the combination of Setting \#1 outperforms all other combinations in all metrics except softness (with $p<0.012$ for these metrics). The advantage of downsampling methods may be because the model fills in the detail in the music sequence instead of modifying it, resulting in higher musical diversity to enhance the listening experiment of audiences. Moreover, the data enhancement methods like duration stretching, noise-making, and transposition may have less to do with the emotion-based music rearrangement task. Nonetheless, we also notice the higher softness of Setting \#2. This may result from its strategy of computing the target emotion as the average value of the last music emotion and the current target emotion, making the emotion-changing curve smoother but decreasing the ability to fit emotions.

}

\begin{table*}[t]
  \small
 \centering
  \caption{{\color{highlight}Objective and subjective comparisons of REMAST and its ablation variants. ($p<0.01$ for all subjective metrics)}.}
 \resizebox{\linewidth}{!}{
 \begin{tabular}{ccccccccccccc}
  \toprule
\multirow{3}{*}{Setting} &\multirow{3}{*}{\makecell{Ablation\\Variants}} & \multicolumn{6}{c}{Objective Metrics} & \multicolumn{5}{c}{{\color{highlight}Subjective Metrics}}  \\
   \cmidrule(lr){3-8}\cmidrule(lr){9-13}
     && \multicolumn{4}{c}{coherence} & \multirow{2}{*}{similarity$\uparrow$} & \multirow{2}{*}{real-time fit$\uparrow$}	& \multirow{2}{*}{coherence}&\multirow{2}{*}{softness}&\multirow{2}{*}{similarity}&\multirow{2}{*}{real-time fit}&\multirow{2}{*}{overall}\\
   \cmidrule(lr){3-6} 
    &  &PCC\cite{yeh2021automatic}$\downarrow$&CEC\cite{yeh2021automatic}$\downarrow$&MCTC\cite{harte2006detecting}$\downarrow$&overall$\uparrow$ & &  &  &  &  &  &  \\
  \midrule
  \#7 & w/o Harmonic Color~ & 3.30±0.28 & 5.37±0.32 &	1.42±0.10  & 4.31±0.43 & 6.41±0.02 & 2.05±0.81  & 2.72 & 2.97 & 2.69 & 2.38 & 10.76 \\

        \#8 & w/o Rhythm Pattern & 3.18±0.28  & 5.63±0.38	& 1.51±0.09  & 4.28±0.47 & 6.52±0.02 & 1.79±0.70  & 2.59 & 2.59 & 2.55 & 2.38 & 10.01 \\ 
 
   \#9 & w/o Contour Factor & 3.50±0.23 & 5.99±0.39 & 1.61±0.14 &  3.90±0.47 & 6.49±0.02 &  2.00±0.74 & 2.55 & 2.76 & 2.48 & 2.24 & 10.03\\ 

 \#10 & w/o Form Factor  & 3.31±0.19	& 6.51±0.30 &  1.61±0.11  & 3.57±0.37  & 6.57±0.03 &  1.99±0.76  & 2.41 & 2.66 & 2.31 & 2.38 & 9.76 \\ 
   \midrule
 \#11 &Bar-level Granularity  &  3.39±0.21  & 6.25±0.35 & 1.40±0.10 &  3.96±0.42  & 6.66±0.03  & 2.08±0.71  & 2.79 & 2.72 & 2.62 & 2.38 & 10.52 \\ 
    \midrule
 \#12 & Semi-supervised Learning &  3.31±0.28  & 5.85±0.48 & 1.55±0.18 &  3.29±0.58  & 5.58±0.14  & \textbf{2.28±0.71}  & 2.72 & 2.55 & 2.03 & 2.34 & 9.66 \\
   \midrule
    & REMAST & \textbf{3.04±0.19} & \textbf{3.71±0.31} &  \textbf{1.04±0.09} & \textbf{6.21±0.37} & \textbf{7.60±0.59} & 2.02±0.74  & \textbf{4.00} & \textbf{3.79} & \textbf{3.90} &\textbf{3.66}& \textbf{15.34} \\ 
        \bottomrule
 \end{tabular}
 }
 \label{tab:ablationresults} 
\end{table*}

\subsection{Ablation Study}

{\color{highlight}
We use four ablation variants of REMAST to investigate the contributions of each music theory feature in real-time emotion-based music arrangement tasks. We remove the respective features and retrain the model. Additionally, we modify REMAST to explore the impact of the granularity and the semi-supervised strategy. All metrics and settings are the same with comparison experiments.
}

Regarding objective metrics in Table \ref{tab:ablationresults}, REMAST outperforms all other ablation variants in overall coherence, similarity, and three coherence sub-metrics (PCC, CEC, and MCTC). This suggests that all ablation variants reduce the acquisition of emotional information from the last timestep's music segment, leading to a decline in music coherence and similarity. Nevertheless, Setting \#7 and Setting \#11 experience some improvements in real-time fit scores. This could be because Setting \#7 reduces the influence of the last timestep's emotional information on the current timestep's music generation by removing the Harmonic Color, significantly increasing the weight of the target input emotion in the emotional dependency of the current timestep's music generation. Setting \#11, on the other hand, fits the target emotion sequences better due to the coarser granularity. However, considering their lower coherence and similarity scores, these variants are not considered viable options. We also discovered that deprecating the semi-supervised strategy to train REMAST on the labeled data (Setting \#12) resulted in the worst performance in other metrics except for real-time fit. The results may be because training on the unlabeled data without emotion recognition loss makes REMAST focus on the music quality while slightly decreasing the emotional accuracy in aligning the generated music and target emotion in objective metrics.

{\color{highlight}
Subjective results in Table \ref{tab:ablationresults} show the effect of each music theory feature, the granularity, and the semi-supervised strategy. We find that deprecating any music features can distinctly decrease the performance of REMAST, which proves that these features play an important role in enhancing the musical and emotional information in REMAST. The results also demonstrate the rationality of using a beat-level granularity. We hypothesize that coarsening granularity to the bar level (Setting \#11) results in a loss of emotional detail information, consequently reducing the coherence of the generated music. The subjective results also prove that even deprecating semi-supervised (Setting \#12) gets slightly high real-time fit scores, it has the worst performance in the subjective experiment, which may result from poor music quality.
}

{
\color{highlight}
\subsection{Evaluation of Emotion Recognition Model}
We evaluated the emotion recognition model and the effect of different music theory features in the emotion recognition model through k-fold validation to prove its robustness and generalize ability. Specifically, we merge datasets EMOPIA and Soundtracks that use discrete emotion representation, then we select each subset of our dataset as the test set while the remaining subsets are used as the train set. We conduct as many experiments as the number of subsets to make sure each datasets are used as the test set. The results are shown in Table \ref{tbl:emo_kfold} that REMAST can generalize to unseen datasets, and all the music theory features can enhance REMAST's emotion recognition ability. The high RMSE in C-WCMED may result from the music genres gap between the Western and Chinese classical music in C-WCMED and modern musical genres in other datasets. We also find that deprecating PMEmo, VGMIDI, EMOPIA, and Soundtracks results in high RMSE on them. The reason may be that these datasets cover various modern music genres and contain a huge quantity of data, enhancing the emotion model's ability. Therefore, the emotion model can obtain low RMSE on DEAM and EmoMusic because their music genres are covered in other numerous training music data.
}

\begin{table*}[htbp]
\centering
\caption{{\color{highlight}Root Mean Square Error (RMSE) of Different Ablated Emotion Recognition Models.}}
\begin{tabular}{lcccccccc}
\toprule
\textbf{Setting} & \textbf{Model} & \textbf{C-WCMED} & \textbf{DEAM} & \textbf{EmoMusic} & \textbf{PMEmo} & \textbf{VGMIDI} & \textbf{EMOPIA+Soundtracks} & \textbf{Average} \\
\midrule
\#13 & w/o Rhythm Pattern & 0.62 & 0.28 & 0.27 & 0.44 & 0.44 & 0.44 & 0.41 \\
\#14 & w/o Contour Factor & 0.61 & 0.30 & 0.28 & 0.42 & 0.43 & 0.45 & 0.42 \\
\#15 & w/o Form Factor & 0.62 & 0.28 & 0.27 & 0.42 & 0.42 & 0.43 & 0.41 \\
\#16 & w/o Harmonic Color & 0.62 & 0.28 & 0.27 & 0.42 & 0.42 & 0.42 & 0.41 \\
\midrule
& REMAST & \textbf{0.59} & \textbf{0.27} & \textbf{0.26} & \textbf{0.42} & \textbf{0.42} & \textbf{0.41} & \textbf{0.39} \\
\bottomrule
\end{tabular}
\label{tbl:emo_kfold}
\end{table*}

\begin{table}[t]
\caption{{\color{highlight}Average Anxiety Relief Degree Comparison($p<0.1$).}}
\begin{tabular}{cccc}
\toprule
\textbf{Metric} & \textbf{REMAST} & \makecell{\textbf{Original}\\\textbf{Songs}}& \makecell{\textbf{Real-time}\\\textbf{Recommendation}}\\
\midrule
\makecell{anxiety relief degree$\uparrow$} & \textbf{21.60} & 12.37 & 4.70 \\ 
\bottomrule
\end{tabular}
\label{tab-regulation}
\end{table}

\begin{figure}
    \centering
    \includegraphics[width=1.0\linewidth]{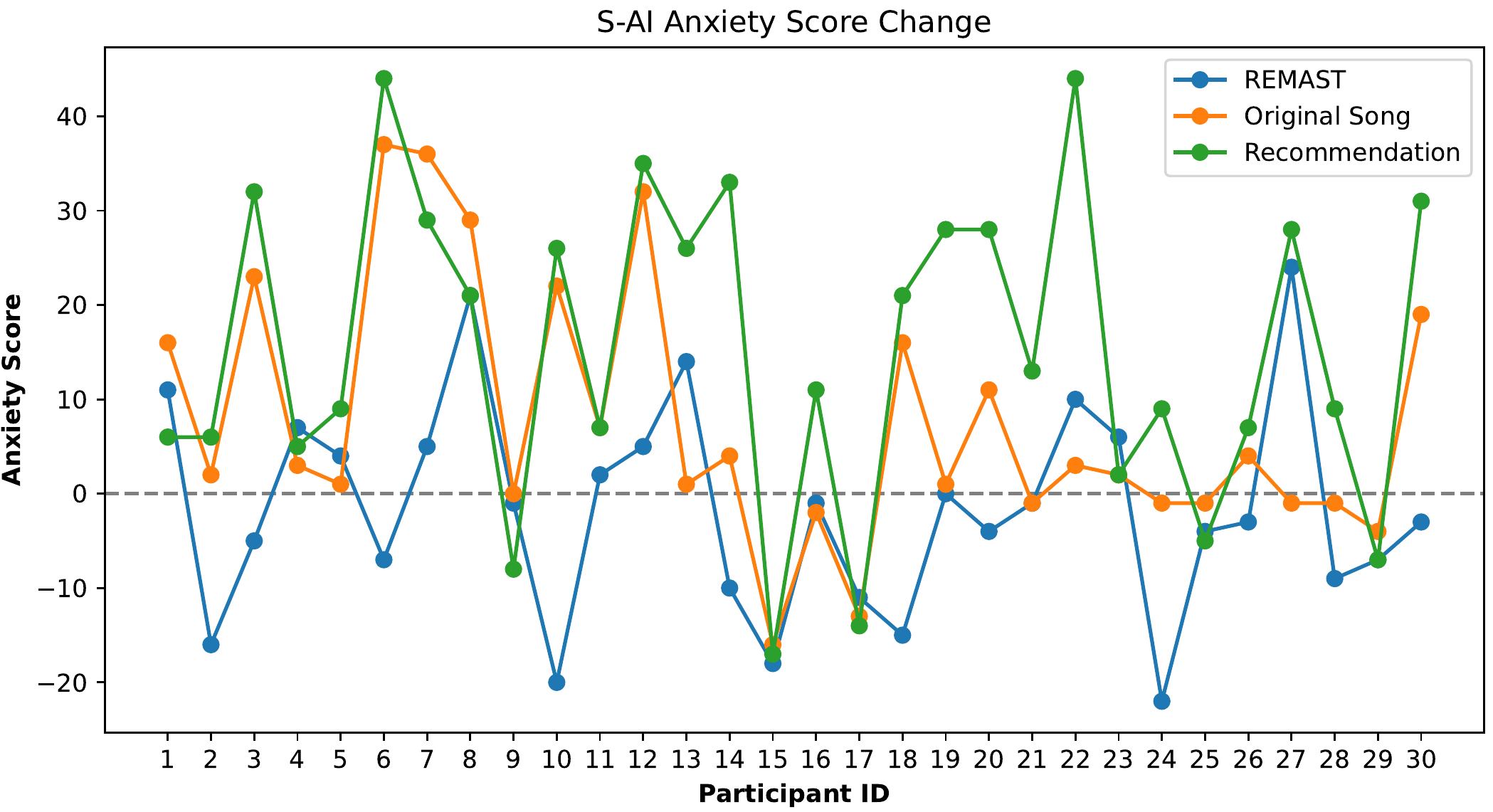}
    \caption{{\color{highlight}The S-AI scores change for each participant of different intervention methods. Points under the horizontal zero line mean participants' anxiety decreases from the state before intervention.}}
    \label{fig:anxiety-relief}
\end{figure}

\begin{table}[t]
\centering
\caption{{\color{highlight}Elapsed Time of REMAST Generating 4-bar Music on Different Computing Devices.}}
\resizebox{0.45\textwidth}{!}{
\begin{tabular}{lc}
\toprule
\textbf{Device} & \textbf{Time (seconds)} \\
\midrule
Intel(R) Xeon(R) Silver 4214R CPU & 2.881 ± 0.037 \\
Apple M1 (MPS) & 2.341 ± 0.083 \\
NVIDIA RTX 3080 TI (CUDA) & 1.015 ± 0.151 \\
\bottomrule
\end{tabular}
}
\label{tbl:gen_time}
\end{table}

{
\color{highlight}
\subsection{Evaluation of Computing Efficiency}
Considering the importance of high computing efficiency in real-time scenarios, we experiment on different devices to obtain the elapsed time of REMAST to generate a four-bar music piece. The results are shown in Table \ref{tbl:gen_time} that REMAST takes nearly three seconds on CPU or one second on GPU to generate a four-bar music piece with a duration of eight seconds when BPM (beats per minute) is 120. Because the elapsed time is less than the music duration, REMAST can fit the real-time scenarios well to generate music based on the target emotion while playing the music to users.
}

\subsection{Application in Anxiety Relief}

{\color{highlight}We further assess emotional regulation in anxiety relief scenarios to validate REMAST's effectiveness in real-world applications. Emotion adjustments in songs by therapists tend to be frequent and have significant guiding effects on participants' emotions during treatment. Applying such emotion sequence to guide the generative model could damage the music coherence. Therefore, a smooth transition plays a crucial role in this scenario.}

Drawing on the RMT (Resource-Oriented Music Therapy) theory\cite{schwabe2005resource} and practical application\cite{KuniyoshiChiko2013RegulativeMusicTherapy}, the therapist on our team created an emotion sequence aimed at guiding participants in alleviating anxiety. The participants' information is the same as described earlier. Before and following each therapy session, participants are asked to complete a State Anxiety Inventory (S-AI)\cite{Spielberger1970ManualFT} to assess the changes in anxiety levels. {\color{highlight}We compute the \textit{anxiety relief degree} by subtracting the difference of pre- and post-S-AI scores from a fixed value. A higher \textit{anxiety relief degree} indicates a higher decrease in anxiety level.}

We use the original, unarranged songs in a comparative experiment to verify the therapeutic benefits added by REMAST. We also introduce emotion-based real-time music recommendation methods\cite{2012Real}\cite{Peng2020A} as a comparison. These methods use the participant's real-time emotion as the target emotion and automatically select music from a database with the same emotion. Using the therapeutic texture generation pattern designed by the music professional in our team, we arrange the ground truth songs to create a therapeutic music database.

{\color{highlight}
The results in Table \ref{tab-regulation} show that REMAST outperforms the original songs and real-time music recommendation methods in anxiety relief ($p<0.1$). We also observed the S-AI score changes before and after the intervention of each method, which is shown in Figure \ref{fig:anxiety-relief}. The results demonstrate that nearly 60\% of the participants' anxiety decreased through REMAST. For the remaining 40\% participants, their anxiety increase level is still very small.This suggests that arranging the original songs to the emotional sequence given by therapists effectively enhances the therapeutic effects of the original songs. Additionally, this demonstrates that real-time music recommendation methods may disrupt the music coherence, leading to participant discomfort due to the music switching. On the other hand, REMAST provides a soft emotional transition in generated music and avoids such discomfort.
}
Through this experiment, we confirm that REMAST has effectiveness in the music therapy application, laying the foundation for its broader applications in various real-world scenarios.

\section{CONCLUSION}
In this study, we propose REMAST, a new method for real-time emotion-based music orchestration, to address the challenge of balancing the degree of emotion fitting as well as the degree of music transition smoothing during the process of intervening with user emotions using music. Compared to other methods, REMAST achieves a good trade-off between real-time fitting of emotions and smooth transition by recognizing the musical emotion of the previous time step and fusing it with the current emotion as a condition for music generation. In addition, through the application of anxiety relief, REMAST offers new possibilities for music therapy applications. In future work, we will enhance REMAST by integrating and analyzing users' electroencephalography (EEG) data for real-time emotional resonance. In addition, we plan to develop and launch a public website that will allow users to interact with REMAST in real time.

\section{Acknowledgement}
This work was supported by the National Natural Science Foundation of China (No.62272409); the Key R\&D Program of Zhejiang Province (No.2022C03126); the Project of Key Laboratory of Intelligent Processing Technology for Digital Music (Zhejiang Conservatory of Music); and the Ministry of Culture and Tourism (No.2022DMKLB001), and the OPPO Research Fund.

\bibliographystyle{ieeetr}
\bibliography{remast}

\vspace*{-\baselineskip}
\begin{IEEEbiography}[{
\includegraphics[width=1in,height =1.25in,clip,keepaspectratio]{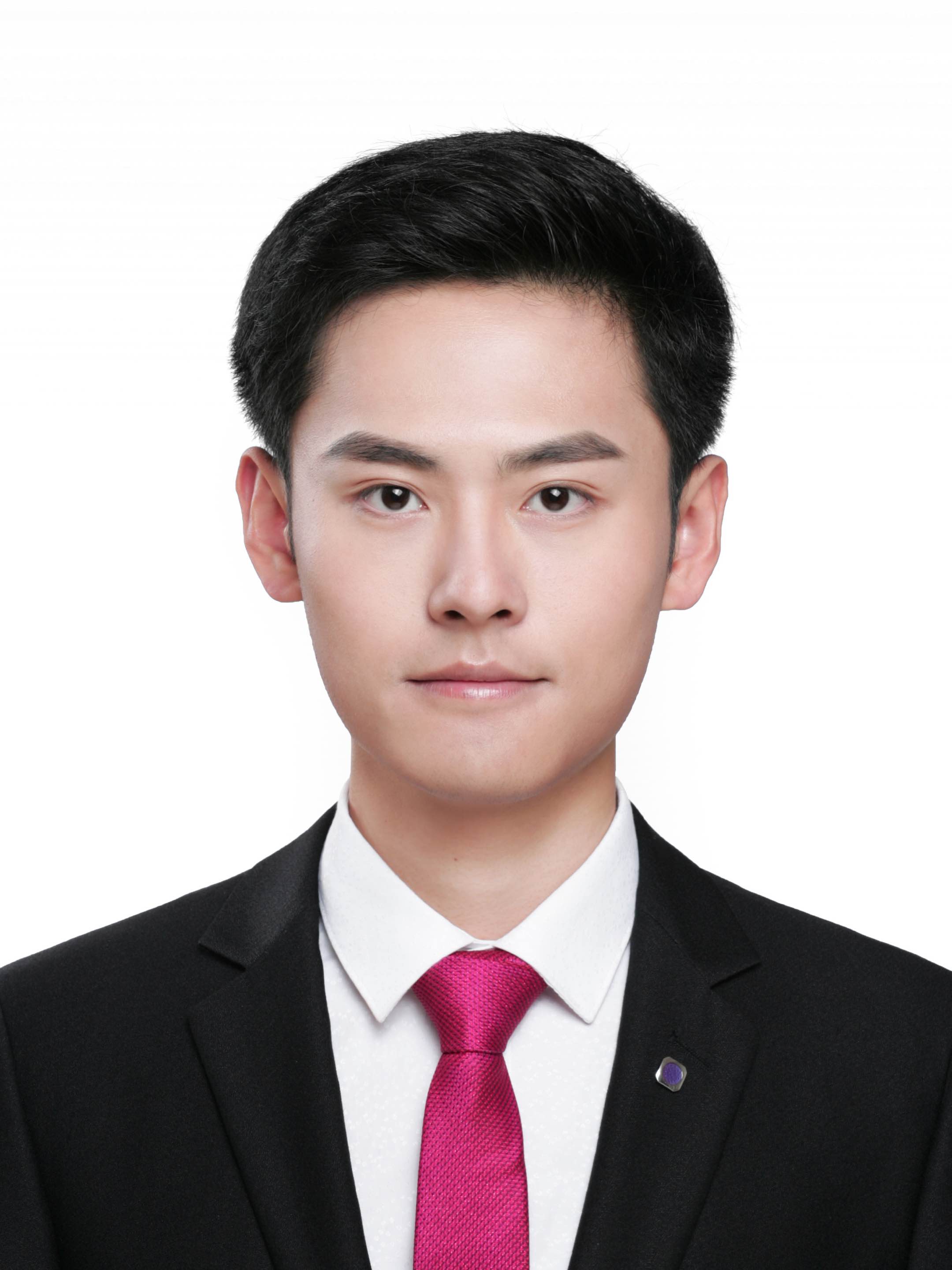}}]{Zihao Wang received his BS degree in Computer Science and Technology from Shandong University in 2021. He is currently pursuing his PhD at the College of Computer Science and Technology, Zhejiang University. His research interests include real-time music generation, emotion-controlled music generation, large-scale audio models, speech and audio processing, as well as the applications of deep learning and machine learning.}
\end{IEEEbiography}
\vspace*{-\baselineskip}

\begin{IEEEbiography}[{
\includegraphics[width=1in,height =1.25in,clip,keepaspectratio]{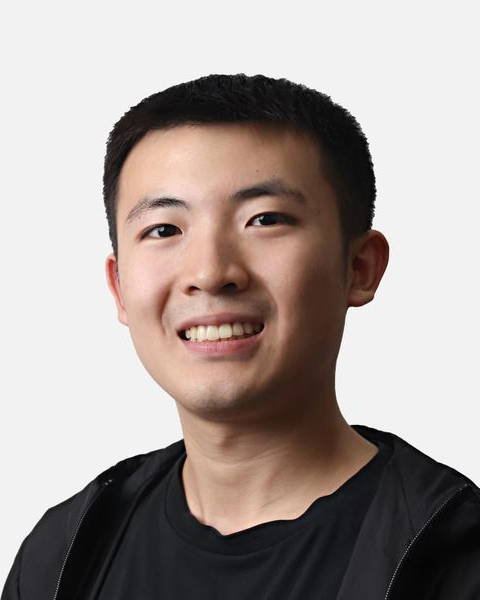}}]{Le Ma received his BS degree in Software Engineering from University of Electronic Science and Technology of China in 2022. He is currently pursuing his PhD at the College of Computer Science and Technology, Zhejiang University. His research interests include speech and audio processing, multi-modal retrieval.
}
\end{IEEEbiography}
\vspace*{-\baselineskip}

\begin{IEEEbiography}[{
\includegraphics[width=1in,height =1.25in,clip,keepaspectratio]{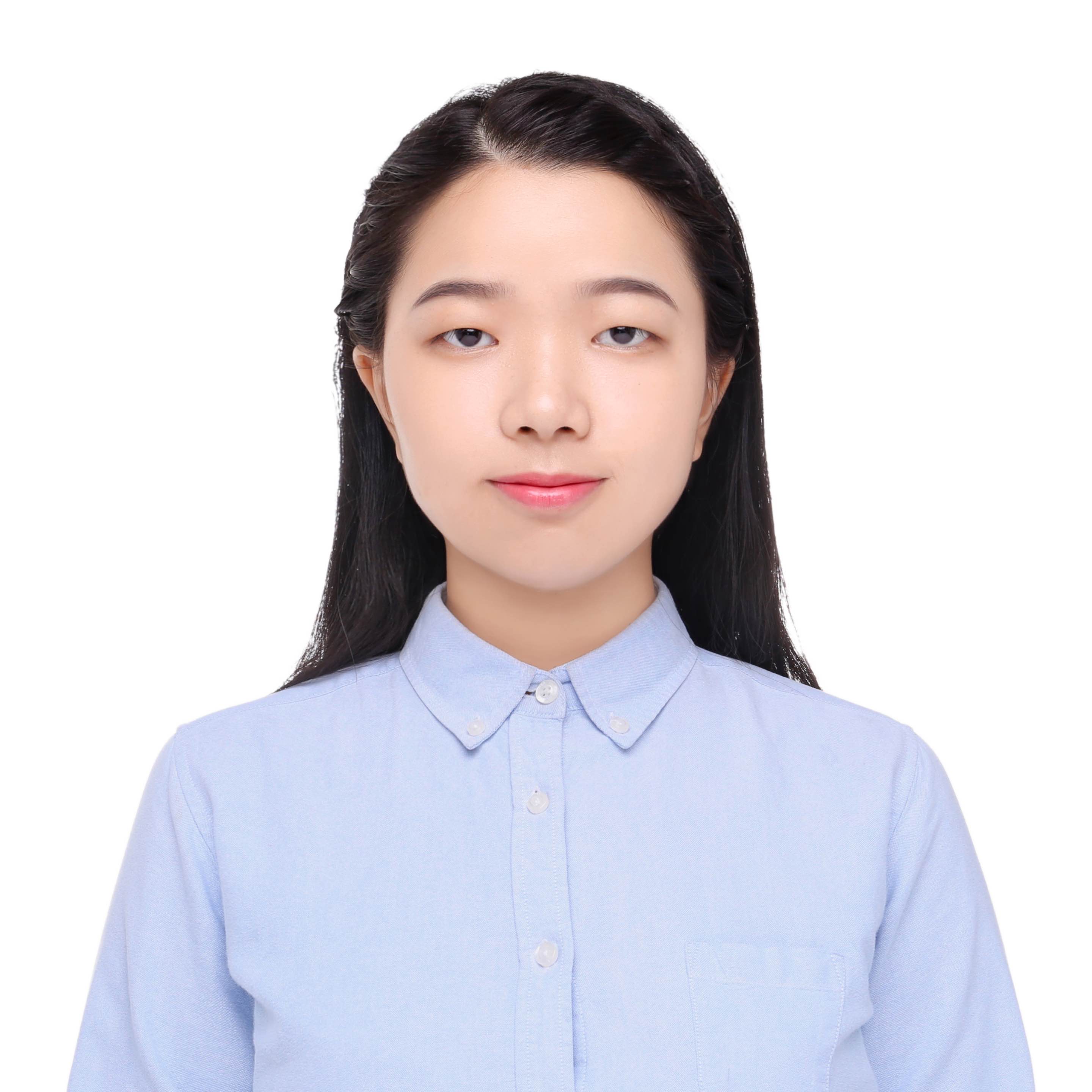}}]{Chen Zhang received the BS degree in digital media technology from Zhejiang University in 2020 and obtained a Master’s degree in Computer Science and Technology
from Zhejiang University in 2023. She is currently a research scientist in ByteDance. Her research interests include speech, AI music, AI avatar and multi-
modal learning.
}
\end{IEEEbiography}
\vspace*{-\baselineskip}

\begin{IEEEbiography}[{
\includegraphics[width=1in,height =1.25in,clip,keepaspectratio]{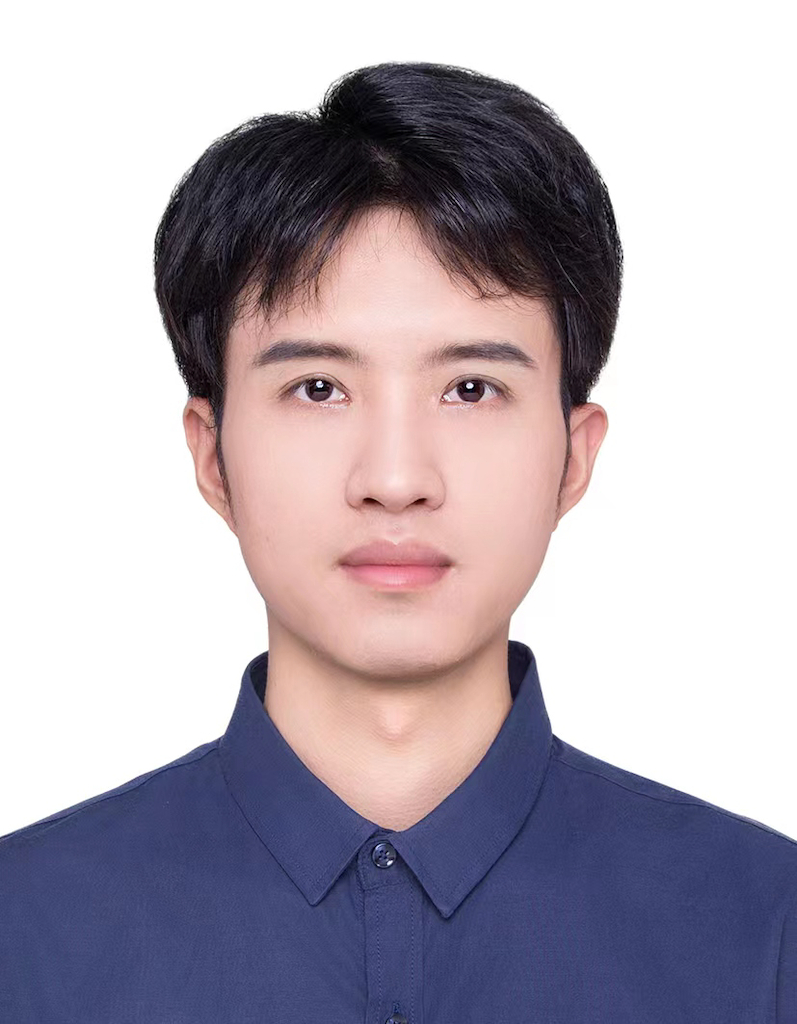}}]{Bo Han is a Ph.D. student in the College of Computer Science \& Technology, Zhejiang University. His primary research interests are in computer vision and multimedia generation.
}
\end{IEEEbiography}
\vspace*{-\baselineskip}

\begin{IEEEbiography}[{
\includegraphics[width=1in,height =1.25in,clip,keepaspectratio]{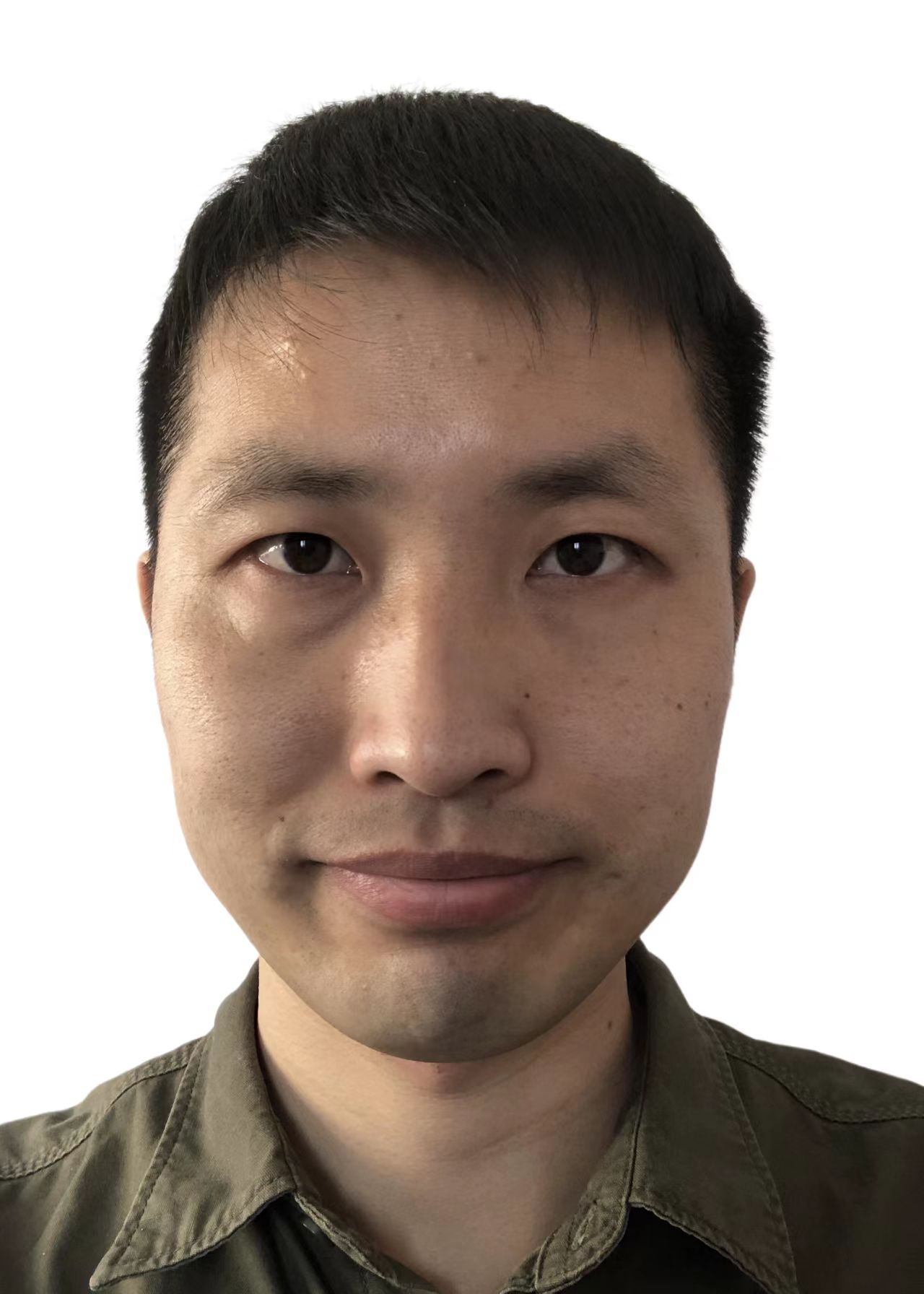}}]{Yunfei Xu received the BS degree in electronic science and technology from Nankai University, Tianjin, China, in 2010, and the PhD degree in signal and information processing from The Institute of Acoustics of The Chinese Academy of Sciences, Beijing, China, in 2015. He is currently an employee of Data \& AI Engineering System, OPPO, Beijing, China. His research interests include speech synthesis, singing voice synthesis, and audio generation.
}
\end{IEEEbiography}
\vspace*{-\baselineskip}

\begin{IEEEbiography}[{
\includegraphics[width=1in,height =1.25in,clip,keepaspectratio]{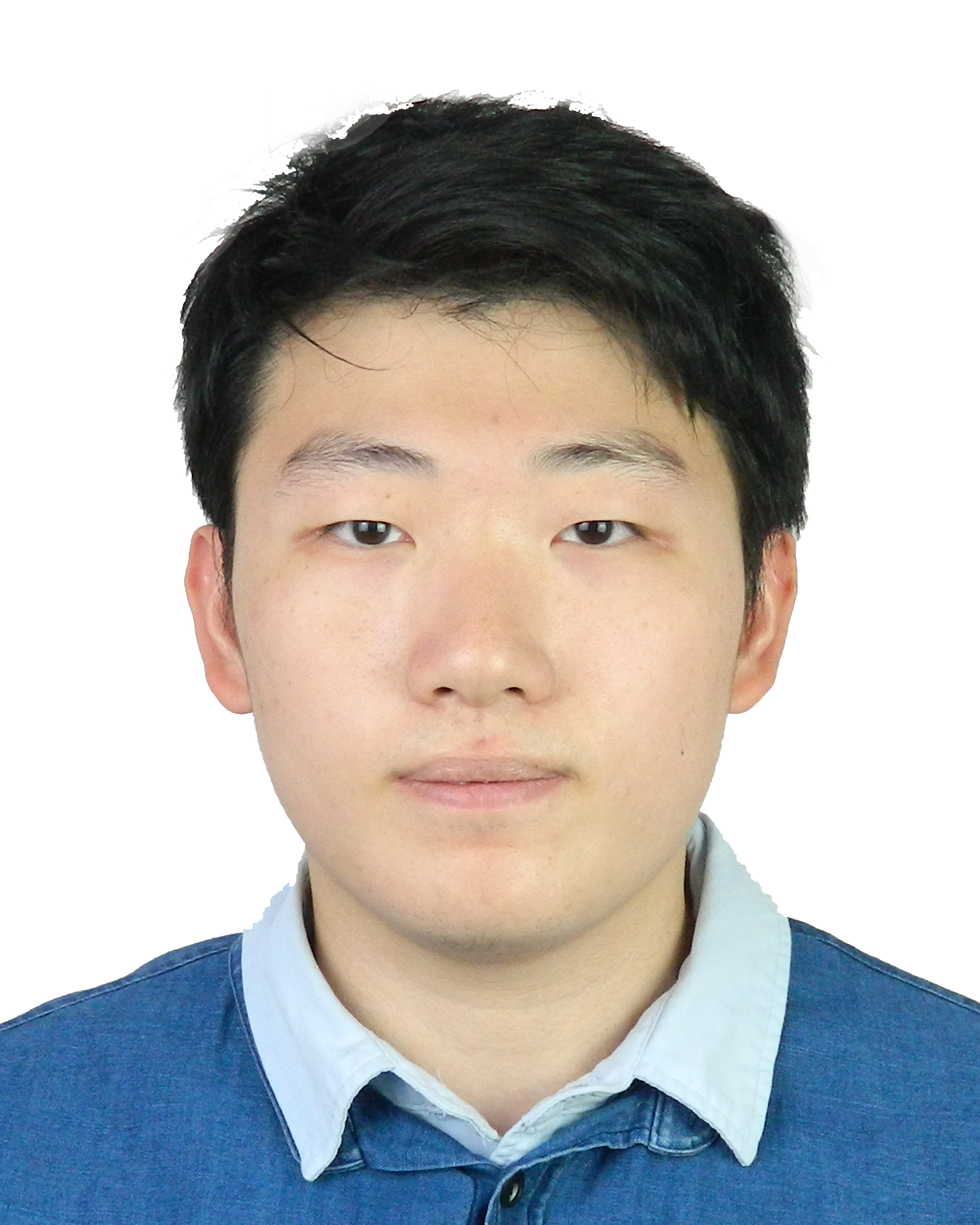}}]{Yikai Wang received the BS degree in computer science from Zhejiang University, China, in 2023. He is currently pursuing PhD degree at the University of North Carolina Chapel Hill. His research interests include reinforcement learning, LLM application, operation research and optimization.
}
\end{IEEEbiography}
\vspace*{-\baselineskip}

\begin{IEEEbiography}[{
\includegraphics[width=1in,height =1.25in,clip,keepaspectratio]{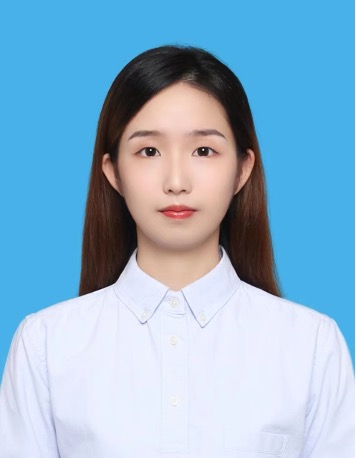}}]{Xinyi Chen is currently working towards the BS degree with the Department of Psychology and Behavioral Science, Zhejiang University, Hangzhou, China. Her research interests include engineering psychology, human-computer interaction, and social computing.
}
\end{IEEEbiography}

\vspace*{-\baselineskip}

\begin{IEEEbiography}[{
\includegraphics[width=1in,height =1.25in,clip,keepaspectratio]{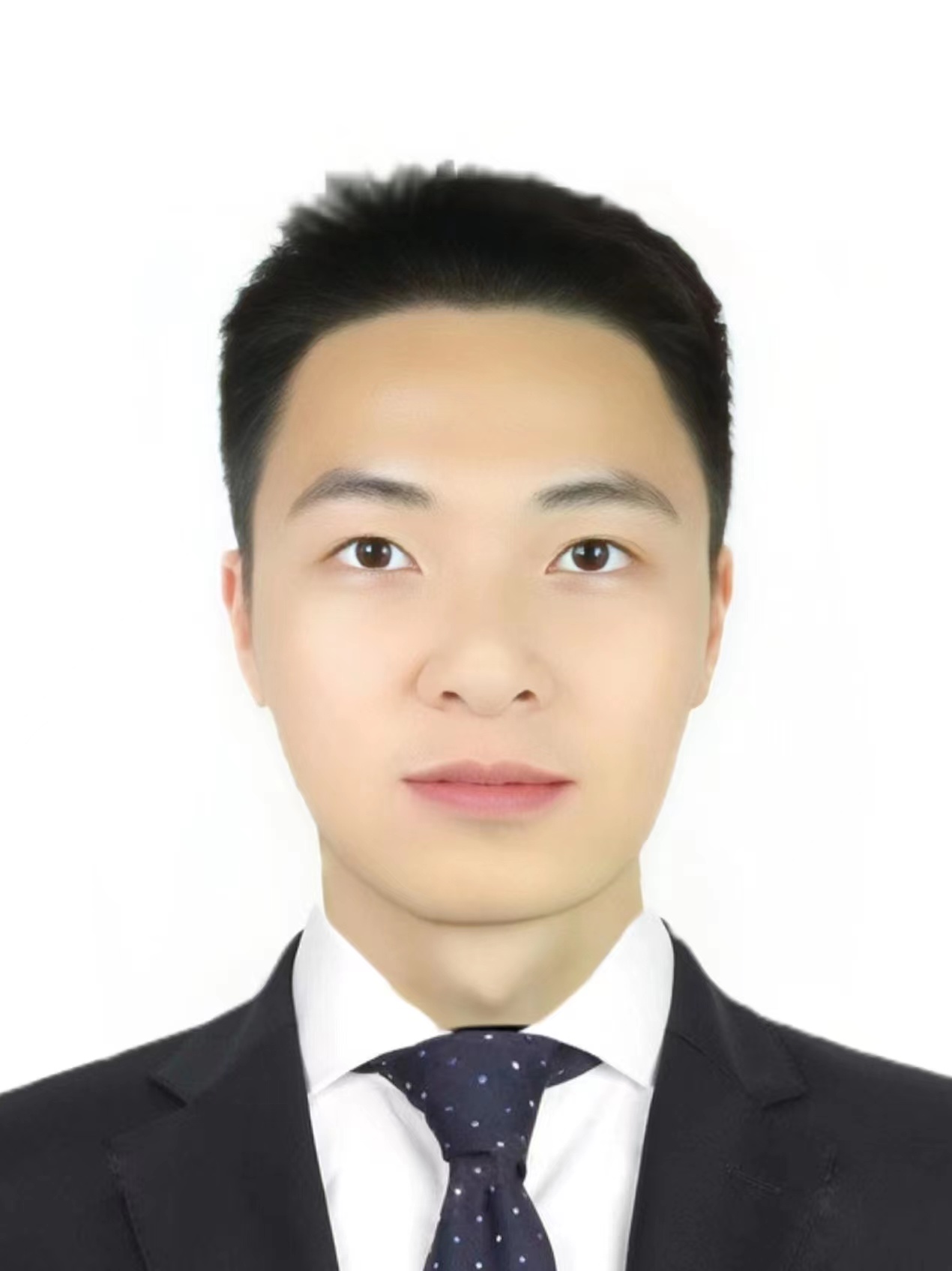}}]{Haorong Hong is currently working towards the BS degree in the college of computer science and technology, Zhejiang University. He has been admitted to Advanced Data \& Signal Processing Laboratory, Peking University and will work for a master's degree there. His research interests include NLP, LLM and LLM fine-tuning.
}
\end{IEEEbiography}

\vspace*{-\baselineskip}

\begin{IEEEbiography}[{
\includegraphics[width=1in,height =1.25in,clip,keepaspectratio]{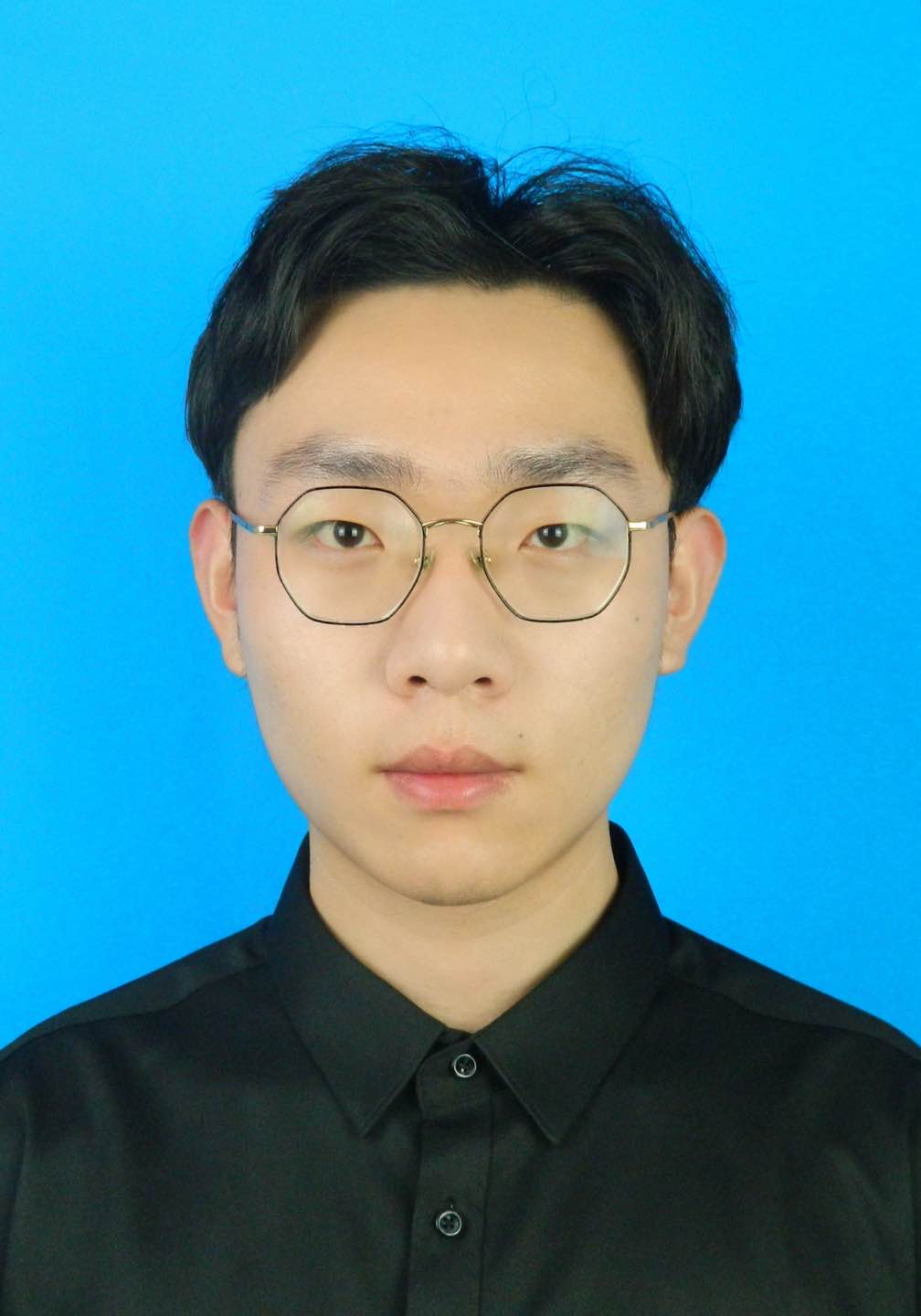}}]{Wenbo Liu received the BS degree in computer science from Zhejiang University, Hangzhou, China, in 2023. He is now pursuing master’s degree in computer engineering at Columbia University, New York, US. His research interests include machine learning, computer vision and IoT.
}
\end{IEEEbiography}
\vspace*{-\baselineskip}

\begin{IEEEbiography}[{
\includegraphics[width=1in,height =1.25in,clip,keepaspectratio]{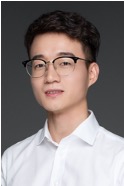}}]{Xinda Wu received the B.Eng. degree from Hangzhou Dianzi University, Hangzhou, China, in 2019. He is currently a Ph. D. student in the College of Computer Science and Technology, Zhejiang University, Hangzhou, China. His research interests include affective computing, human-computer interaction, multimedia information retrieval, and intelligent music generation.
}
\end{IEEEbiography}
\vspace{-\baselineskip}

\begin{IEEEbiography}[{
\includegraphics[width=1in,height =1.25in,clip,keepaspectratio]{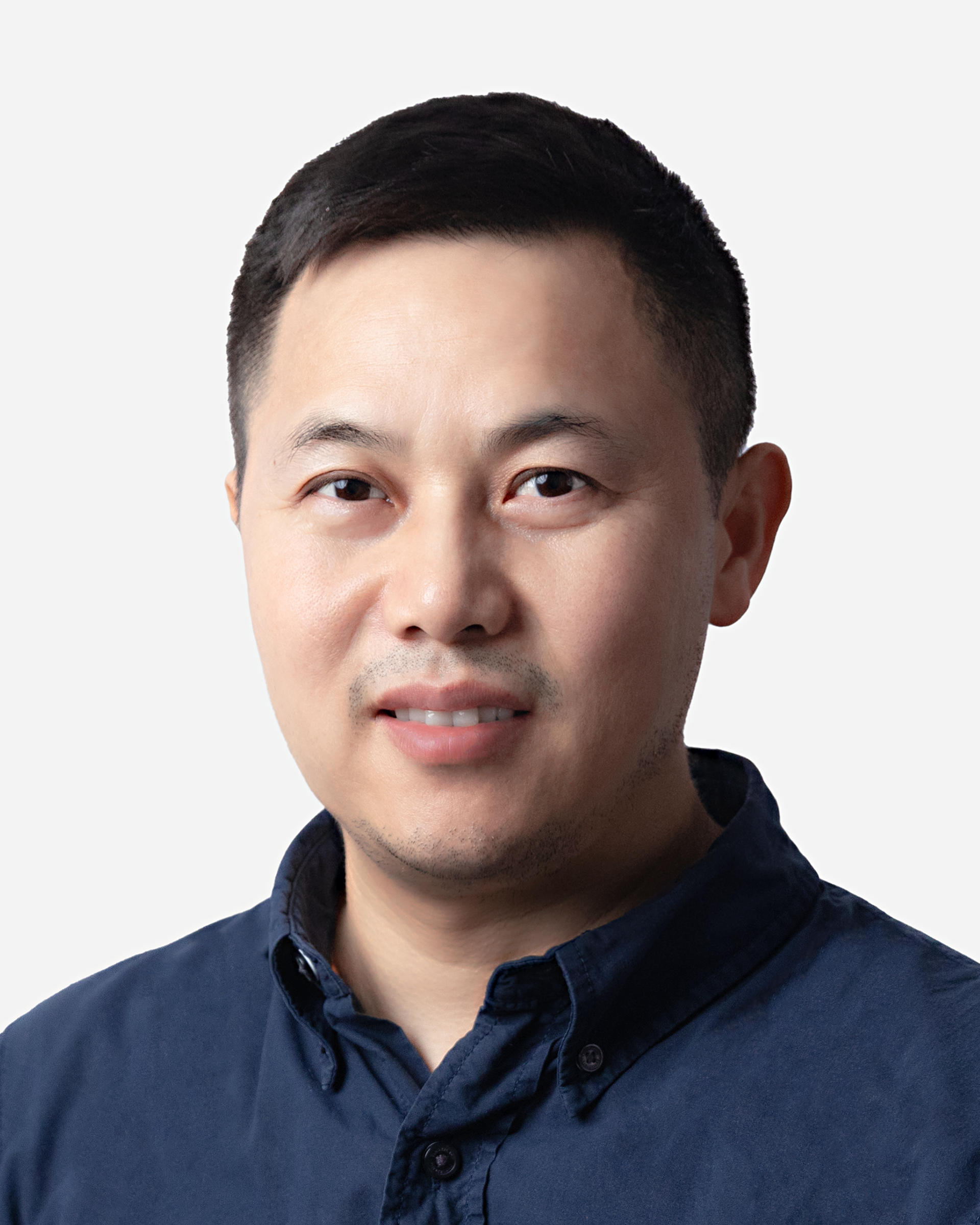}}]{Kejun Zhang (Member, IEEE) received the PhD degree in computer science from Zhejiang University, China. He is currently a professor with the College of Computer Science, Zhejiang University, China. His research interests include music information retrieval, artiﬁcial intelligence and machine learning. He has published many research papers in various reputable journals and conference proceedings.
}
\end{IEEEbiography}
\vspace{39.5\baselineskip}

\end{document}